\documentclass[twocolumn,amsmath,amssymb,floatfix]{revtex4-1}

\usepackage{dcolumn}
\usepackage{bm}
\usepackage{graphicx}
\usepackage{epstopdf}
\usepackage{xcolor}
\usepackage{wrapfig}

\usepackage{natbib}

\def\kT{k_{{}_{\rm B}}T}
\def\kappabar{\bar\kappa}

\newcommand{\argmin}{\operatornamewithlimits{arg\,min}}

\newcommand{\argmax}{\operatornamewithlimits{arg\,max}}

\begin{document}

\title{Solvated Membrane Nanodiscoids: A Probe For The Effects Of Gaussian Curvature
}

\author{R.~G.~Morris${}^{1,2,3}$, T.~R.~Dafforn${}^{4}$ and 
M.~S.~Turner${}^{1}$}
\affiliation{${}^1$Department of Physics and Centre for Complexity Science, 
University of Warwick, Coventry CV4 7AL, UK}
\affiliation{${}^2$Simons Centre for the Study of Living Machines, National 
Centre for Biological Sciences, TIFR, 
Bangalore, 560064, India}
\affiliation{${}^3$EMBL-Australia, University of New South Wales, Sydney, Australia}
\affiliation{${}^4$School of Biosciences, University of Birmingham, Birmingham, 
B15 2TT, UK}

\date{\today}
%

	%
	%
\begin{abstract}
	Several methods now exist to solvate lipid bilayer discoids at the scale of tens of nanometres.  Due to their size, such nanodiscoids have a comparatively large boundary-to-area ratio, making them unusually well-suited to probing the effects of Gaussian curvature. Arguing that fluctuations in discoid size and shape are quenched on formation, we quantify the stability, in terms of size and shape, of near-solvation discoid-like flaps that are subject to thermal fluctuations.  Using cryo-Electron Microscopy images of Styrene Maleic Acid stabilised discoids, we deduce that stable, saddle-like discoids (with high Gaussian curvature) can likely be solvated from bulk lamellar ($L_\alpha$) phase at moderate-to-high surface tensions ($>10^{-4}$ N/m).  We then describe how such tension-controlled solvation can be used for both measuring, and fractionating membrane components according-to, the modulus of Gaussian rigidity $\kappabar$.  Opportunities for investigating the effects of Gaussian curvature on membrane-embedded proteins, which can be co-solvated during the formation process, are also discussed.
\end{abstract}

\maketitle

\section{Introduction}
\begin{figure}[!t]
	\begin{center}
		\includegraphics[width=0.45\textwidth]{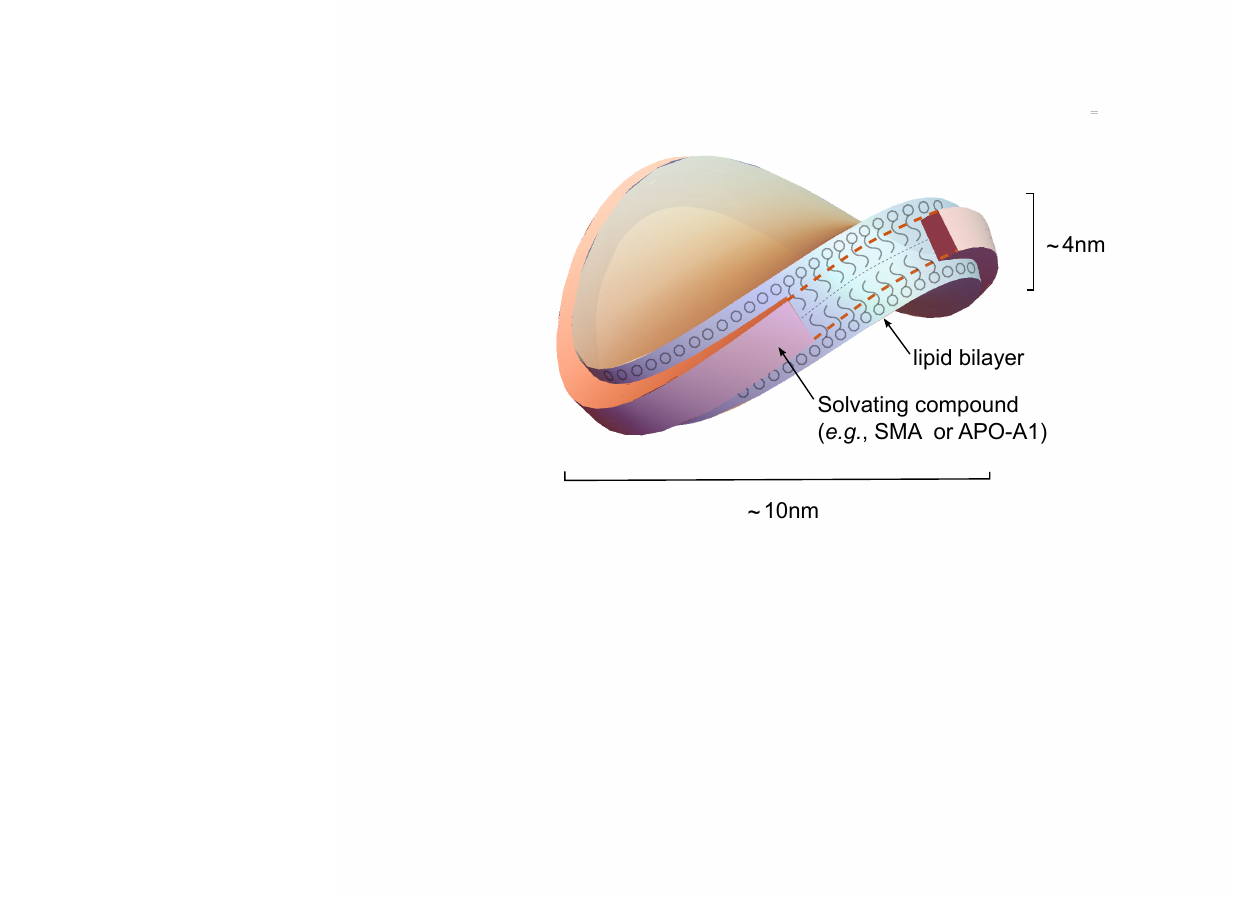}
	\end{center}
	\caption
	{
		(Colour online) Sketch of a saddle-shaped nanoscale lipid bilayer discoid.  The structure is stabilised by the amphipathic nature of the solvating compound ({\it e.g.} SMA or APO-A1): the hydrophobic groups interact with the acyl chains of the lipids, whilst the hydrophilic groups face the solvent.  The large boundary-to-area ratio of these structures makes them ideally suited for probing phenomena that couple to Gaussian curvature.
}
\label{fig:car}
\end{figure}

Several long-chain molecules have now been shown to solvate nanoscale lipid bilayer discoids, with two prominent examples being the copolymer Styrene Maleic Acid (SMA) \cite{Knowles2009,Tonge2000,Tonge2001,Lee2016} and the $\alpha$-helical lipoprotein APO-A1 \cite{Bayburt2002,Schuler2013,Denisov2016}. The key feature of these molecules is their amphipathic structure: in the presence of a lipid bilayer, the hydrophobic groups interact with the acyl chains of the lipids, whilst the hydrophilic parts face the solvent, resulting in a solution of stable nanodiscoids (see Fig.~\ref{fig:car}).  Significantly, such discoids have been shown to preserve the integrity of co-solvated transmembrane proteins, implying an important role for the purification, structural determination, and functional characterization of membrane proteins 
\cite{Orwick-Rydmark2012,Paulin2014,Dorr2014,Gulati2014,Postis2015,Denisov2016,Sun2018,Qiu2018a}.

The modulus of Gaussian rigidity, $\kappabar$, quantifies a membrane's preference for forming locally saddle-like or spherical shapes, and determines the equilibrium topology of bulk membrane phases, {\it e.g.}, a highly connected {sponge} ($L_3$), one or more (large) membranes ($L_\alpha$) or a large number of {small vesicles} \cite{Safran1994}.  However,  $\kappabar$ is  notoriously difficult to measure. This is because of the Gauss-Bonnet theorem \cite{Helfrich1973,Safran1994,DoCarmo1976}, which states that Gaussian curvature doesn't couple to any observable deformation modes for one component membranes of fixed, boundary-free topology.  Even for membranes with a boundary (or a hole \cite{Turner2004}), the coupling to Gaussian curvature is usually negligible for all but the largest boundary-to-area ratios.  Solvated nanodiscoids have an extremely large boundary-to-area ratio, making them not only technologically important, but also uniquely suited for studying the role of $\kappabar$ in membranes.

\begin{figure*}[!t]
	\begin{center}
		\includegraphics[width=0.98\textwidth]{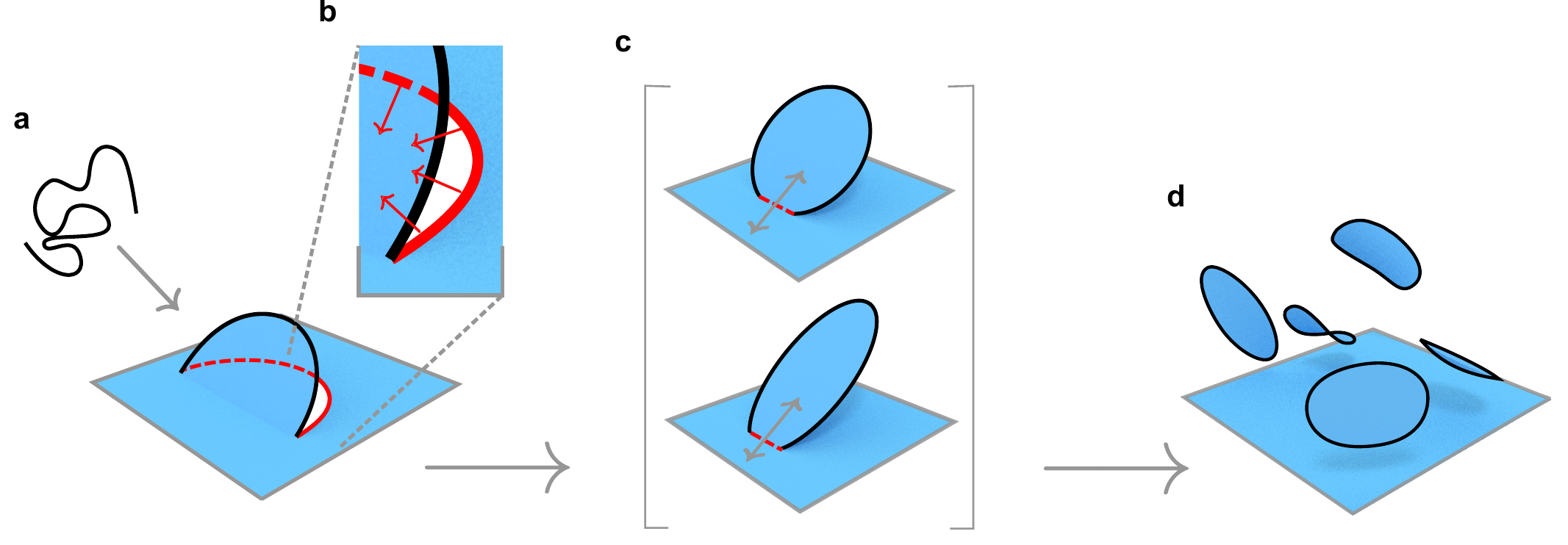}
	\end{center}
	\caption
	{
		(Color online) Cartoon of the putative formation process.  Driven by its amphiphilic character, the solvating long-chain molecule inserts into the membrane ({\bf a}).  The presence of the solvating compound leads to a hole and a free edge (red dashed line) in the membrane.  The high energetic cost of such free edges, however, drives the closure of such holes, first forming a straight `cut' and then bringing the two ends of the solvating molecule together ({\bf b}).  Near solvation, a small membrane flap remains ({\bf b} and {\bf c}) which can still exchange molecules with the bulk membrane.  On `pinching-off', the differences in flap size and shape that result from such fluctuations become quenched, since the elastic modulus of membranes held at fixed particle number is extremely large. The result is a polydisperse mixture of solvated nanodiscoids ({\bf d}).}
\label{fig:mix}
\end{figure*}
In this article we study how lipid nanodiscoids can be used to probe phenomena that couple to Gaussian curvature.  We show how they can be used to both indirectly measure $\kappabar$, and fractionate lipids by $\kappabar$ into discoids with negative Gaussian curvature.  More generally, given the extreme values of Gaussian curvature, both positive and negative, found in vesicles and organelles associated with protein synthesis and sorting \cite{BA+89,Nixon-Abell2016}, it is natural to ask: what are the effects of this curvature on membrane-embedded proteins, and does it play a role in sorting.  In this context, lipid nanodiscoids may have a wider role as a tool for investigating the effects of Gaussian curvature on co-solvated transmembrane proteins.

\section{Polydispersity Quenched On Formation}
We begin with the observation that solvated nano-scale discoids form a relatively polydisperse mixture, with an apparently broad range of areas and shapes.  Since lipid bilayers are essentially incompressible, we conclude that such variation is a result of the formation process, which is poorly understood.  Moreover, we note that the timescales of dissipation, due to either the viscosity of the bilayer, or the surrounding water, are comparatively short.  We therefore assume a quasi-static description of nanodisc solvation, which we use to quantify the statistics of area and shape variation, and hence imply the aforementioned new role for nanodiscoids in the context of the  probing of Gaussian curvature.  

Neglecting any potential complications associated with adsorption/insertion, our starting point is a fully-inserted protein/co-polymer restricted to planar self-avoiding conformations ({\it i.e.}, crossings are not allowed).  It is assumed that the (known) amphipathic properties of the long-chain solvating molecule then lead the formation of a flap of bilayer, able to protrude discontinuously from the membrane (see Fig.~\ref{fig:mix}).  The large energy-per-unit-length associated with an unadorned bilayer edge then ensures that the ``hole'' left behind by such a flap quickly becomes a ``cut'', which effectively runs in a straight line between the two ends of the solvating protein/co-polymer (see red and red-dotted lines in panels {\bf a}, {\bf b} and {\bf c} of Fig.~\ref{fig:mix}).  Further energy minimization then drives the meeting of the two free ends, reducing the cut length, and leading to discoid solvation.  Very close to formation, when the cut is very small, an effective line tension can be associated with the closed curve comprising the boundary of the near-solvation flap and the cut itself.

Provided that it is very near the point of ``closure'' (when the protein/co-polymer ends meet) and that it retains the structure (and symmetry) of a standard bilayer, a Helfrich-like continuum theory \cite{Helfrich1973} can be used for the bulk of the flap.  Assuming uniform lipid composition, the leading order energetics \cite{DeGennes1993} are related to the geometry of the midsurface $\mathcal{S}$ of the discoid-like flap via
\begin{equation}
	\mathcal{H}_m = \int_\mathcal{S} dA
	\left[
		\sigma + \frac{\kappa}{2}\left( 2H - c_0\right)^2 + \bar{\kappa} K 
	\right],
	\label{eq:H_m}
\end{equation}
where $\sigma$ is a surface tension, set by exchange of lipids through the almost-closed gap between protein/co-polymer ends.  As usual, $H$ and $K$ are the mean and Gaussian curvatures, and $\kappa$ is the membrane bending rigidity.  We assume that there is no density mismatch between bilayer leaflets and therefore take $c_0 = 0$ throughout.

\begin{figure*}[!t]
	\vspace{3mm}
	\begin{center}
		\includegraphics[width=0.75\textwidth]{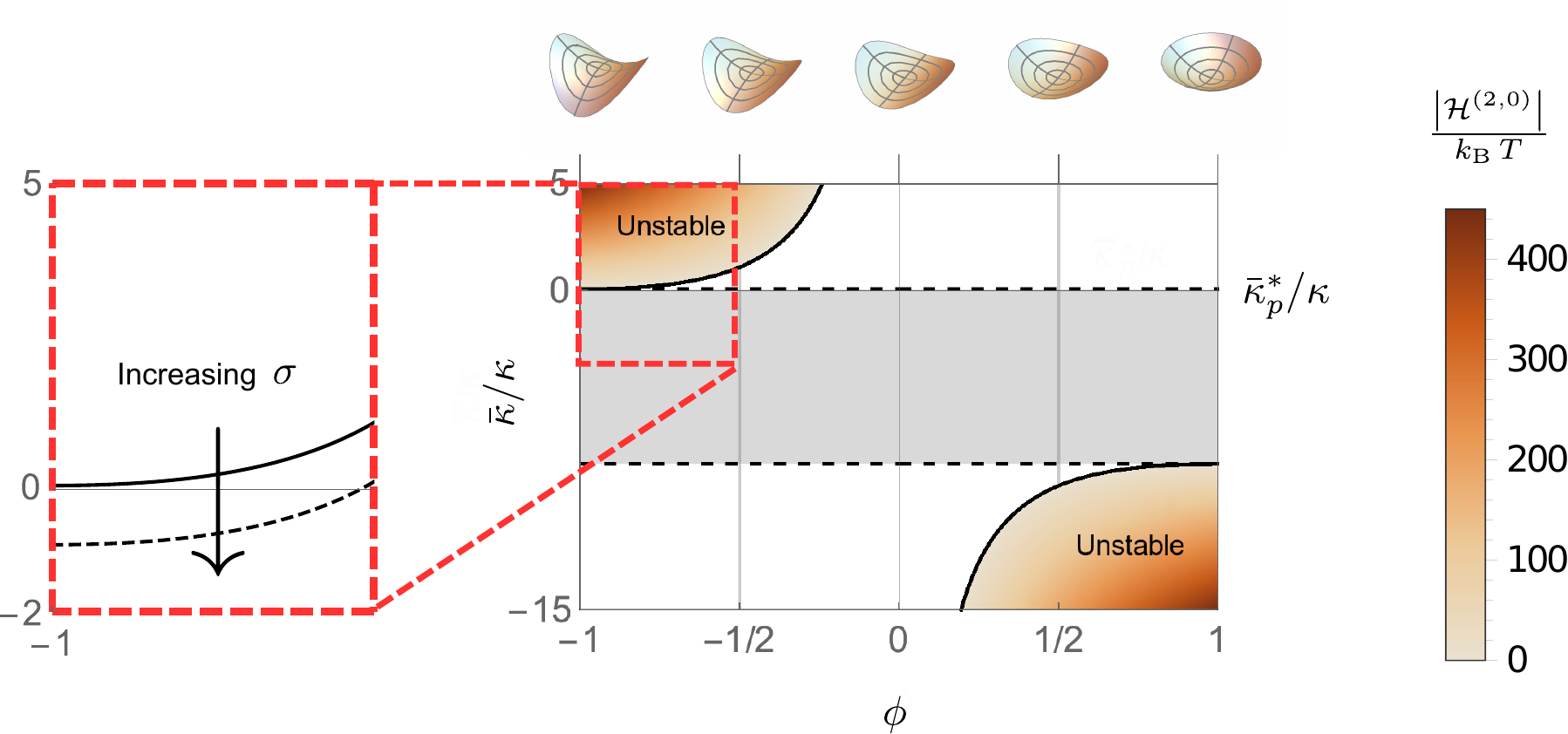}
	\end{center}
	\caption
	{
		(Color online) Linear stability of out-of-plane perturbations $\phi$, plotted against the ratio of bending moduli $\kappabar/\kappa$, for indicative values $\kappa = 20\,\kT$, $k_g = k_n = R_0^3\,\sigma$, $R_0 = 5$ nm, $\sigma = 10^{-4}$ N m$^{-1}$, and $q_0 = 1/R_0$.  Regions of stability/instability are delimited by solid black lines, which correspond to the quantity $\kappa^\ast(\phi)/\kappa$ (see SM).  Within a band of values, all perturbations are suppressed and only flat discs are stable (grey shaded region).  Outside of this band, a range of shapes are unstable, with pringles (saddles with principal curvatures of equal magnitude) or shallow spherical caps being dominant when $\kappabar/\kappa$ is above or below the stable region, respectively (see `heatmap' shading, which quantifies the growth rate of a given perturbation).  The critical value $\kappabar^\ast_p = \kappabar^\ast(\phi=-1)$ therefore controls the transition to pringles, which spontaneously form only if $\kappabar^\ast_p$ is less than or equal to the Gaussian ridgidity of the bilayer in question ($\kappabar=-\kappa$ for a wide variety of lipids \cite{MH_JJB_MD_12}). From Eq.~(\ref{eq:kappabar_p}), we see that $\kappabar^\ast_p$ decreases linearly with $\sigma$, indicating that with sufficient tension pringles can always be rendered unstable, irrespective of the composition of the bilayer used, and hence its modulus of Gaussian rigidity (see inset).		}
\label{fig:stab}
\end{figure*}
There is also an energy associated with boundary, $\partial\mathcal{S}$, which 
is the closed line formed by concatenating the solvating protein/co-polymer and 
the remaining small cut.  We include both a line tension $\tau$, and an 
energetic cost to bending (amphipathic asymmetry precludes any twist).  By 
symmetry, the latter must be invariant under sign change of both the normal and 
geodesic curvatures of $\partial\mathcal{S}$, for which we use the symbols 
$q_n$ and $q_g$, respectively.  The contribution is written as the following 
line integral
\begin{equation}
	\mathcal{H}_b = \int_{\partial\mathcal{S}} dl \left[ \tau + 
	\frac{k_g}{2}\left( q_g - q_0 \right)^2 + \frac{k_n}{2}\,q_n^2 \right],
	\label{eq:H_e}
\end{equation}
where $k_g$ and $k_n$ are effective bending moduli.  The former corresponds to 
in-tangent-plane bending, whilst the latter corresponds to out-of-tangent-plane 
bending. [For definitions of the relevant geometrical quantities, see the 
Supplementary Material (SM)].  Notice that since the solvating 
protein/co-polymer is assumed to preserve the symmetry of the bilayer on 
reflection about $\mathcal{S}$ (true for both SMA and APO-A1), a non-zero 
spontaneous line curvature ($q_0$) is only permitted in the tangent plane of 
the surface, and not in the normal direction.

Of course, once solvated, lipid exchange with the membrane (reservoir) is no 
longer possible.  All subsequent area and shape fluctuations are then 
suppressed by the disproportionately large intrinsic elastic modulus of lipid 
bilayers (when held at fixed particle number) and, as a result, any 
thermally-induced fluctuations in area due to lipid exchange when close to solvation become fixed, or ``quenched'', for all time at the point when the discoid finally detaches.   
\begin{figure*}[!t]
	\begin{center}
		\includegraphics[width=0.9\textwidth]{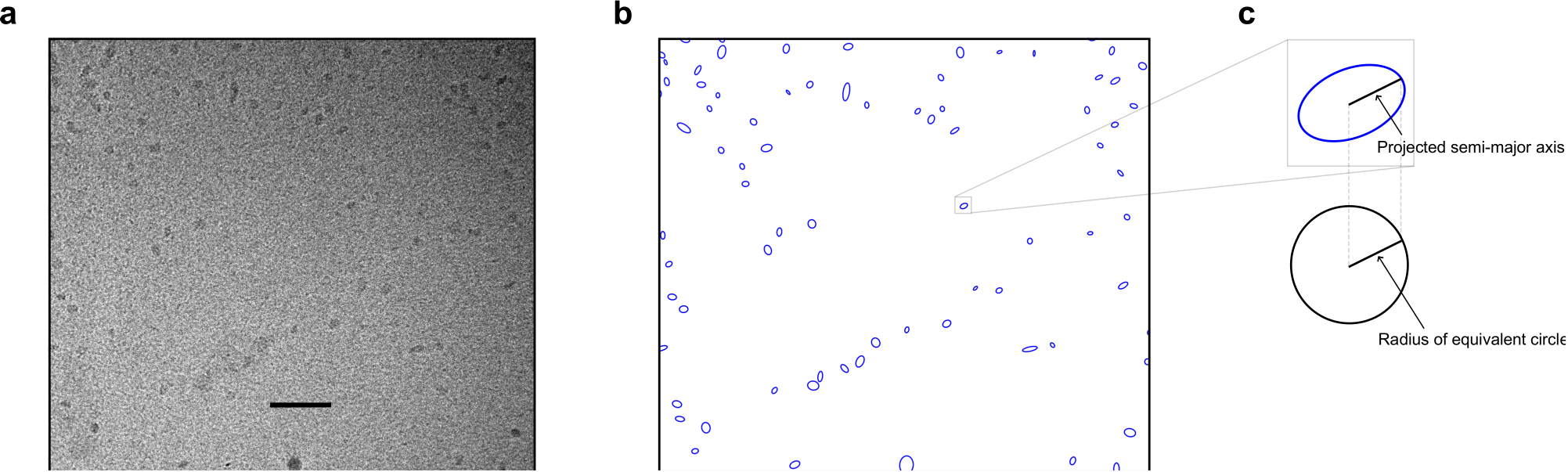}
	\end{center}
	\caption
	{
		(Color online).  Cryo-Electron Microscopy image ({\bf a}) of SMA 
		stabilised discoids, prepared according to \cite{Lee2016} (scale bar 
		$=50$ nm).  Sufficiently large values of $k_g$ and $k_n$ correspond to 
		isotropically oriented mixtures of flat discs, whose radial 
		fluctuations are governed by Eq.~(\ref{eq:MSD}) with $n=0$.  Therefore, 
		using image processing tools to fit ellipsoids to each discoid ({\bf 
		b}) and extracting the mean-squared deviation of the semi-major axes 
		({\bf c}), Eq.~(\ref{eq:MSD}) may be used to calculate an upper 
		estimate of $k_g \simeq 7\,\kT\,\mathrm{nm}$ and therefore an upper 
		bound on $\sigma_p(\kappabar)$, the tension above which pringles form.
}
\label{fig:im}
\end{figure*}

\section{Tension Controlled Instability to Pringle Shapes}
To understand the ramifications of these ideas, consider flat shapes ({\it 
i.e.}, $H=K=q_n=0$).  In this case, the total energy, $\mathcal{H} = 
\mathcal{H}_b + \mathcal{H}_m$, is minimised by discs of radius $R_0$, which is 
related to the line tension $\tau$ (and $k_g$, $q_0$ and $\sigma$) via 
\begin{equation}
	\tau = \frac{k_g}{2R_0^2}\left( 1 - q_0^2\,R_0^2\right) - R_0\,\sigma,
	\label{eq:tau}
\end{equation}
%
(see SM for details).  The in-plane stability can be investigated by writing an 
angle-dependent radius $R(\theta) = R_0 + \delta R (\theta)$ such that $\delta 
R (\theta) = \epsilon\,R_0\,\sum_n \Re\left[A_n\, \exp\left( i\,n\, 
\theta\right) \right]$, where $\Re$ indicates the real part, $A_n \in 
\mathbb{C},\ \vert A_n\vert \le 1$ for all $n\in\mathbb{N}$, and $\epsilon \ll 
1$ is dimensionless.  The energy, $\mathcal{H}$, can then be expanded as a 
power series in terms of $\epsilon$.  Here, the mean-squared thermal amplitude 
of the $n^{\rm th}$ fluctuation mode follows by the principle of equipartition 
of energy (see SM) \cite{Chaikin1995}
\begin{equation}
	\epsilon^2\, \left\langle \left\vert A_n \right\vert^2\right\rangle
	= \frac{\kT\,R_0}{\pi\left( n^2 - 1 \right) \left[ k_g \left( n^2 - 1 
	\right) - R_0^3\,\sigma \right]}.
	\label{eq:MSD}
\end{equation}
The $n=0$ mode is a dilation, and is always stable (we expect $\sigma,k_g>0$ 
for all lipid discs, irrespective of whether they are cut from living cells, 
adsorbed bilayers or vesicles).  The $n=1$ mode corresponds to a translation 
where the disc remains circular to lowest order, and is of indeterminate 
amplitude.  In general, for $n>1$, a mode of degree $n$ is stable if $k_g \ge 
R_0^3\,\sigma/(n^2 -1)$.  This puts a lower bound $k_g \ge R_0^3\,\sigma/3$ on 
the geodesic rigidity of the attached solvating compound that is required for 
stable circular discs.

For non-flat discoids, we consider the subset of all out-of-plane deformations 
that can be parametrised by the two (orthogonal) principal curvatures at the 
centre, $c_1$ and $c_2$.  This is reasonable provided that the disc is 
(laterally) much smaller than the membrane correlation length 
$\sqrt{\kappa/\sigma}$, which is typically at least a few tens of nm.  The 
curvatures are labelled, without loss of generality, such that $\vert c_1\vert 
\ge \vert c_2\vert$.  Using a polar Monge approach, the height field is chosen 
to be
\begin{equation}
	h(r,\theta) = \frac{\alpha\,r^2}{4\,R_0}\left[ 1 + \phi + \left(1-\phi 
		\right) \cos 2 \theta
	\right],
	\label{eq:h_alpha}
\end{equation}
where the angle $\theta$ is measured from the direction of $c_1$, $\phi = 
c_2/c_1$ takes values in the interval $[-1,+1]$, and $\alpha = R_0\,c_1 \ll 1$ 
is a small dimensionless parameter.  Both mean and Gaussian curvatures remain 
constant across the disc up to $O\left( \alpha^2 \right)$ making (\ref{eq:H_m}) 
straightforward to calculate (see SM).  However, this is not the case for the 
either line element at the boundary $dl$ or the line curvatures $q_g$ and $q_n$ 
appearing in (\ref{eq:H_e}).  Indeed, both $dl$ and $q_g$ also rely on 
$\epsilon$, and we therefore adopt a notation where the $O\left( 
\alpha^a\,\epsilon^e \right)$ term in an expansion of $\mathcal{H}$ is written 
as $\alpha^a\,\epsilon^e\,\mathcal{H}^{(a,e)}$.  The resulting expression (see 
SM) has neither any first order terms [{\it i.e.}, at $O(\alpha)$ or 
$O(\epsilon)$], nor any $O(\alpha\,\epsilon)$ cross terms:
\begin{equation}
	\mathcal{H} = \mathcal{H}^{(0,0)} + \epsilon^2\, \mathcal{H}^{(0,2)} + 
	\alpha^2\,\mathcal{H}^{(2,0)} + O\left(\epsilon^3\right) + O\left( \alpha^3 
	\right).
	\label{eq:H_general}
\end{equation}
The coefficients $\mathcal{H}^{(0,0)}$ and $\mathcal{H}^{(0,2)}$ can therefore 
be read-off from the previously calculated energy of a flat disc [$\tau$ is 
unchanged from (\ref{eq:tau})].  Moreover, the earlier stability analysis holds 
too, but now refers to the {\it projected} disc size, which is independent of 
out-of-plane perturbations to orders $\epsilon^2$ and $\alpha^2$.

The stability of out-of-plane perturbations is governed by the sign of 
$\mathcal{H}^{(2,0)}$, which depends not only on $\phi$, but also the material 
properties of the solvating polymer ($R_0$, $k_g$, $k_n$, and $q_0$), the 
membrane bending moduli ($\kappa$ and $\kappabar$), and the membrane tension 
$\sigma$, which is set during formation.  We focus on how the properties of the 
membrane ({\it i.e.,} tension and bending) affect discoid stability and, in 
particular, the role of $\kappabar$.  Leaving the details to the SM, we write 
$\mathcal{H}^{(2,0)}=\pi\,\phi\,\left[ \kappabar - \kappabar^\ast (\phi) 
\right]$, which introduces a critical value of Gaussian rigidity, 
$\kappabar^\ast$.  Fig.~\ref{fig:stab} shows how this quantity determines the 
stability of a given perturbation $\phi$, as a function of the ratio $\kappabar 
/ \kappa$.  Also shown is the magnitude of $\mathcal{H}^{(2,0)}$, and hence the 
rate at which the perturbation grows or shrinks in time.  For a band of values 
of $\kappabar / \kappa$, all perturbations are suppressed and only flat discs 
are stable.  Outside of this band, perturbations characterised by a range of 
$\phi$ are unstable.  However, the dominant ({\it i.e.}, fastest growing) mode 
always corresponds to either $\phi=-1$ or $\phi=+1$.  That is, a saddle with 
principal curvatures of equal magnitude--- a ``pringle''--- or a shallow 
section of a sphere, respectively.  In the case of the former, the instability 
that takes a flat disc to a pringle is just $\kappabar \ge \kappabar^\ast_p$, 
where the critical value 
\begin{equation}
	\kappabar^\ast_p = \left(\frac{3}{2\,R_0} -q_0\right)k_g + 
	\frac{k_n}{2\,R_0} -\frac{R_0^2\,\sigma}{4},
	\label{eq:kappabar_p}
\end{equation}
is marked in Fig.~\ref{fig:stab}.  For a given $\kappabar$ therefore, pringles 
are always unstable for tensions above some critical value $\sigma_p$, which 
can be found by inverting (\ref{eq:kappabar_p}).

We can assess whether such tensions $\sigma\ge\sigma_p(\kappabar)$ are 
physically plausible by analysing cryo-Electron Microscopy images of SMA 
stabilised discoids.  When combined with equipartition (\ref{eq:MSD}), such 
images permit the calculation of an effective upper bound on $\sigma_p$ by 
estimating the maximum possible values of $k_g$ and $k_n$.  Figure \ref{fig:im} 
shows one such cryo-EM image, prepared according to \cite{Lee2016}--- {\it 
i.e.}, stabilised by SMA and cut from moderate tension vesicles 
($\sigma\sim$10$^{-4}$ N/m) that were synthesised from {\it E.~Coli} lipid 
extract ($\kappabar \simeq -\kappa$ \cite{MH_JJB_MD_12}).  For such systems, if 
$k_g$ is sufficiently large to stabilise the fluctuations in projected discoid 
shape and size ($\ge R_0^3\,\sigma/3$), then out of plane fluctuations are also 
suppressed if $k_n \ge k_g$.  (That is, the line $\bar{\kappa}=-\kappa$ is 
within the grey region of Fig.~\ref{fig:stab}).  In this case, equilibrium 
discoids are flat discs, and any variation in shape appearing in 
Fig.~\ref{fig:im} is a result of projection onto the focal plane.  Using image 
processing, each discoid may be fitted to an ellipse, and the mean-squared 
deviation of the semi-major axes may be calculated (Fig.~\ref{fig:im} and SM).  
The result can be used to invert Eq.~(\ref{eq:MSD}) at $n=0$ and predict an 
upper estimate for the geodesic rigidity of $k_g = 7\,\kT\,\mathrm{nm}$. 
Further assuming $k_g= k_n$ and setting $q_0= 1/R_0$, the aforementioned upper 
estimate for $k_g$ yields an {\it upper} bound on $\sigma_p$ of $\sim 10^{-2}$ 
N/m, a tension at which bilayer membranes typically rupture \footnote{By using 
	three-dimensional imaging, for example, a better (lower) estimate of $k_g$ 
	should be possible, leading to an estimate of $\sigma_p$ which is below the 
tension at which bilayers tear}.  We deduce, therefore, that pringle-shaped 
discoids can likely be cut from $L_\alpha$ phase membrane at large, but 
physically plausible, tensions.  Moreover, by considering higher order 
contributions to the energy, it can be shown that the resultant pringles are 
stable for moduli of Gaussian rigidity $\kappabar 
\geq\mathrm{max}\left(\kappabar_p^\ast,\kappabar_p^\dagger\right)$ (see SM), 
where $\kappabar_p^{\dagger}$ a critical value similar to that of 
Eq.~(\ref{eq:kappabar_p}), but which is determined at $O\left( \alpha^4 
\right)$.

\section{Fractionation According to Modulus of Gaussian Curvature}
Significantly, the pringle transition can occur {\it before} the bulk membrane 
phase switches from lamellar ($L_\alpha$) to sponge ($L_3$) \cite{Safran1994}.  
That is, saddle-shaped discoids can form spontaneously after being cut from 
flat sheets, as opposed to being cut from a bilayer with equilibrium negative 
Gaussian curvature, such as the sponge phase.  Stable pringles are interesting because they promote the 
preferential sorting of lipids with larger intrinsic Gaussian rigidities into 
the discoids as they are formed.  To illustrate this we consider a simple 
two-component model in which an SMA stabilised pringle emerges from a 
well-mixed (bulk) membrane having an average Gaussian rigidity $\kappabar$ in 
the regime $\kappabar 
>\mathrm{max}\left(\kappabar_p^\ast,\kappabar_p^\dagger\right)$.  Relative to 
the bulk membrane, the pringle may contain an additional area fraction $\psi$ 
of one of the components, resulting in a Gaussian rigidity for the pringle of 
the form $\kappabar+\delta\kappabar\,\psi$, with 
$\delta\kappabar=\kappabar_1-\kappabar_2>0$ the difference between the Gaussian 
rigidity of each component. The membrane Hamiltonian might then include an 
extra term
\begin{equation}
	\mathcal{H}_\psi = \int_\mathcal{S}\frac{\chi\,\psi^2}{2}dA,
	\label{eq:extra}
\end{equation}
with $\chi$ a Flory parameter that approaches zero as the bulk membrane 
approaches the demixing transition.  The inclusion of (\ref{eq:extra}) implies 
that the disc spontaneously adopts the shape of a stable pringle, with 
curvature given by
\begin{equation}
	\left\vert c_1\right\vert = \frac{1}{R_0}\left[\frac{2\left( 
	\kappabar-\kappabar_p^\ast \right)}{3\left( \kappabar - 
	\kappabar^{\dagger}_\chi \right)}\right]^{1/2},
	\label{eq:c_1}
\end{equation}
when 
$\kappabar>\mathrm{max}\left(\kappabar_p^\ast,\kappabar_\chi^\dagger\right)$ 
for $\kappabar^{\dagger}_\chi = \kappabar^{\dagger}_p - 
2(\delta\kappabar)^2/\chi\,R_0^2$ (see SM).
The corresponding equilibrium composition of the discoid is characterised by 
the additional area fraction 
\begin{equation}
	\psi = \frac{2\,\delta\kappabar\,\left( \kappabar - \kappabar_p^\ast 
	\right)}{3\,\chi\,R_0^2\,\left( \kappabar - \kappabar^{\dagger}_\chi 
\right)},
	\label{eq:psi}
\end{equation}
which is occupied by lipids with modulus of Gaussian rigidity $\kappabar_1$.  

\section{Conclusions and Discussion}
Taken as a whole, our analysis suggests the possibility of using lipid nanodiscoids to not only phase-separate, but actually {\it fractionate} membrane components by $\kappabar$; those components with the largest values of $\kappabar$ will be preferentially sorted into pringles. It is also possible that the instability described here could be used as the basis of a technique to measure or compare values of $\kappabar$. One possible approach is to slowly  increase the surface tension of vesicles via micropipette-aspiration or osmotic control: saddle shapes should first form when $\kappabar=\kappabar^\ast_p (\sigma)$.  We re-iterate that these results are especially perinent given the lack of available alternatives that prob lipid coupling to membrane Gaussian curvature. 

We propose that such ideas can readily be extended to areas of significant Biological interest: the localisation or activities of membrane-bound proteins may also depend on Gaussian curvature via both the protein's effective shape and elastic response to deformation.  Indeed, given that membrane-bound proteins can be co-solvated with their integrity preserved, the nanoscale discoids described here provide an interesting new quantitative technology for studying the wider role of Gaussian curvature in cell Biology--- {\it e.g.}, the sorting and function of membrane-bound proteins in organelles whose function is closely tied to morphology, such as the Endoplasmic Reticulum or Golgi Apparatus.  We therefore welcome further work in the area.

\subsection*{Acknowledgments}
R.G.M and M.S.T acknowledge the support of EPSRC grant \# EP/E501311/1, a 
	Leadership fellowship to M.S.T.  R.G.M also acknowledges the Simons 
	Foundation and the Tata Institute of Fundamental Research.  T.R.D 
	acknowledges the support of a BBSRC BRIC project grant \# BB/G010412/1.

\subsection*{Author contributions}
M.S.T and T.R.D conceived-of and designed the research.  R.G.M and M.S.T 
performed analysis and T.R.D contributed images/data.  R.G.M and M.S.T wrote 
the manuscript in consultation with T.R.D.


\clearpage

\begin{widetext}

\section*{Supplementary Material}

\section*{Introduction}
The purpose of this supplementary material is threefold.  First, to provide any 
necessary theoretical background (in a terse, but self-contained way), 
particularly with respect to differential geometry (for further details, the 
reader is referred to \cite{DoCarmo1976,frankel}).  Second, to provide a detailed 
account (making heavy use of the aforementioned background) of the calculations 
whose results appear in the main manuscript.  Lastly, to describe the protocol 
used to analyse the cryo Electron Microscopy images.

\section*{Geometry}
\subsection*{Setup and notation}\label{sec:setup}
Let $\mathcal{S}$ be a smooth Reimannian manifold representing the nanodiscoid 
midsurface.  If points on the midsurface are labelled by an ``internal'' 
coordinate $u\in\mathbb{R}^2$, then $\mathcal{S}$ is just the image of $u$ 
under an embedding $F:\mathbb{R}^2\to \mathbb{R}^3$.  That is, the position in 
$\mathbb{R}^3$ of a point $u$ is given by the vector $\boldsymbol{R}(u) =: 
F(u)$.

In terms of notation, the convention, which will be used throughout, is that 
Greek indices take values 1 or 2, whilst Latin indices take 1, 2 or 3.  Bold 
typeface is used to represent vectors in $\mathbb{R}^3$, such that 
$\boldsymbol{v} = v^i\,\hat{\boldsymbol{e}}_i$, where an implicit sum is 
understood by repeated indices of different type ({\it i.e.}, upper and lower), 
and $\{\hat{\boldsymbol{e}}_i:i=1,2,3\}$ are the usual normalised basis 
vectors, independent of position $\boldsymbol{R}(u)$.  By contrast, an 
overarrow is used for vectors in $\mathbb{R}^2$, such that $\vec{v} = 
v^\alpha\,\vec{e}_\alpha$, with $\{\vec{e}_\alpha:\alpha=1,2\}$ representing 
non-normalised basis vectors, which depend on position $u$. 

The embedding $F$ corresponds to a pushforward $F_{\ast}:T_u\mathbb{R}^2\to 
T_{\boldsymbol{R}(u)}\mathbb{R}^3$, whose operation is defined by:  
\begin{equation}
	\boldsymbol{F}_\ast\left(\vec{e}_\alpha\right) = \frac{\partial 
		\boldsymbol{R}\left(u^\alpha\right)}{\partial u^\alpha}.
	\label{eq:pos}
\end{equation}
The notion of $\boldsymbol{F}_\ast\left( \vec{e}_\alpha \right)$ can be used to 
define the unit normal to $\mathcal{S}$:
\begin{equation}
	\hat{\boldsymbol{n}} = \frac{\boldsymbol{F}_\ast\left( \vec{e}_1 \right) 
	\times \boldsymbol{F}_\ast\left( \vec{e}_2 \right)}{\left\vert 
		\boldsymbol{F}_\ast\left( \vec{e}_1 \right) \times 
		\boldsymbol{F}_\ast\left( \vec{e}_2 \right)\right\vert},
	\label{eq:nhat}
\end{equation}
where the symbol $\times$ represents the usual cross product in $\mathbb{R}^3$.  
(For the purposes of this article, it suffices to assume that the orientation 
of the basis of a manifold is the same as the orientation of the manifold 
itself).

\subsection*{Forms}\label{sec:forms}
A form $r^{(k)}$ of degree $k$ is a linear functional that takes $k$ tangent 
vectors and returns a real number.  For example, a 0-form is a function, a 
1-form is a co-vector, and higher forms of degree $k$ are tensors of rank 
$(0,k)$.  (For notation, the degree of a form will be indicated by a bracketed 
superscript unless explicitly stated otherwise).

\subsubsection*{1-forms}
At a point $u$, the basis of 1-forms that spans the cotangent space, 
$T^\ast_u\mathbb{R}^2$, is given by $\{\mathrm{d}u^\alpha:\alpha=1,2\}$, such 
that $\mathrm{d}u^\alpha (\vec{e}_\beta) = \delta^\alpha_\beta$, where 
$\delta^\alpha_\beta$ is the Kronecker delta symbol.  In this basis, $p = 
p_\alpha \mathrm{d}u^\alpha$, where once again there is an implicit sum over 
repeated indices.

For example, consider a curve $\mathcal{C}$, parameterised by $t$ and embedded 
in $\mathcal{S}$.  $\mathcal{C}$ has tangent vectors $\boldsymbol{T} = 
\partial\boldsymbol{R}/\partial t$ with pre-image $\vec{T}$ under $F_\ast$.  In 
this case, the line element ``$dl$'' is just
\begin{equation}
	dl:	=\left\Vert \,\vec{T}\, \right\Vert\,\mathrm{d}t
	= \left[ g_{\alpha\beta} T^\alpha
		\,T^\beta
	\right]^{1/2}\,\mathrm{d}t.
	\label{eq:dl_def}
\end{equation}
Equivalently, ``$dl$'' can also be defined relative to the Euclidean 
coordinates of the embedding space:
\begin{equation}
	dl 
	= \left\{ \delta_{ij} \left[\boldsymbol{F}_\ast 
	\left(\vec{T}\right)\right]^i
	\left[\boldsymbol{F}_\ast \left(\vec{T}\right)\right]^j
	\right\}^{1/2}\,\mathrm{d}t,
	\label{eq:dl_def_2}
\end{equation}
where the implied sum is now over $i,j=1,2,3$ and not $\alpha,\beta=1,2$ as in 
(\ref{eq:dl_def}) and $\delta_{ij}$ is just the metric in Euclidean 
coordinates.

\subsubsection*{First fundamental form, metric and inner product}\label{sec:met}
The embedding function $F$ induces a metric on $\mathbb{R}^2$ via the pullback 
$F^\ast$.  That is $g_{\alpha\beta}(u) = \left( F^\ast\,\mathrm{I} 
\right)\left( \vec{e}_\alpha,\vec{e}_\beta \right) = 
\boldsymbol{F}_\ast\left(\vec{e}_\alpha 
\right)\cdot\boldsymbol{F}_\ast\left(\vec{e}_\beta \right)$, where $\mathrm{I}$ 
is the first fundamental form of $\mathbb{R}^3$ ({\it i.e.}, with coefficients 
$\delta^{ij}$) and ``$\cdot$'' is the usual dot product.  At each point 
$u\in\mathbb{R}^2$, the induced metric can be used to define an inner product 
$\langle \cdot,\cdot\rangle:T_u\mathbb{R}^2\times T_u\mathbb{R}^2\to 
\mathbb{R}$.  That is, for arbitrary vectors $\vec{v}$ and $\vec{w}$, we define 
$\langle\vec{v},\vec{w}\rangle :=
v^\alpha g_{\alpha\beta} w^\beta$.  Such an inner product permits the explicit 
identification of a vector, {\it e.g.}, $\vec{v}$, with its dual 1-form, $v$, 
by the condition $v(\vec{w})=\langle \vec{v},\vec{w}\rangle$, which holds for 
all $\vec{w}$.  Noticing that $v(\vec{w}) = v_\alpha \mathrm{d}u^\alpha 
(\vec{w}) = v_\alpha w^\alpha$ and using the above definition of the inner 
product of two vectors implies the raising and lowering properties of the 
metric and its inverse [$g^{\alpha\beta} = \left( g_{\alpha\beta} 
\right)^{-1}$], respectively.  That is, $v_\alpha = g_{\alpha\beta} v^\beta$ 
and $v^\alpha = g^{\alpha\beta} v_\beta$.  Using this property, the inner 
product acting on two 1-forms can be defined in a complementary way to that of 
the inner product on vectors:
\begin{equation}
	\langle v,w\rangle := v_\alpha g^{\alpha\beta} w_\beta = v_\alpha w^\alpha= 
	\langle\vec{v},\vec{w}\rangle.
	\label{eq:inner_1_form}
\end{equation}

\subsection*{Covariant derivative}\label{sec:covar}
The covariant derivative at a point $u$ on $\mathrm{S}$ is a generalisation of 
the directional derivative.  It calculates the rate of change of a tensor field 
(at $u$) whilst moving along the unique geodesic that has tangent vector with 
pre-image $\vec{y}$ (at $u$) under $F_\ast$.

\subsubsection*{Scalars}
The action of the covariant derivative on a scalar field $\phi$ is defined to 
be
\begin{equation}
	\nabla_{\vec{y}} \,\phi := \phi_{,\alpha}\,\mathrm{d}u^\alpha (\vec{y}),
	\label{eq:cov_scalar}
\end{equation}
where a subscript comma ``,'' is shorthand for a partial derivative, {\it 
i.e.}, ${\phi}_{,\alpha} := \partial \phi / \partial u^\alpha$.

\subsubsection*{Vectors}
When acting on a tangent vector $\vec{v} = v^\alpha \vec{e}_\alpha$, we write
\begin{equation}
	\nabla_{\vec{y}} \,\vec{v} := \vec{e}_\alpha \left( 
	{v^\alpha}_{;\beta}\right)\mathrm{d}u^\beta (\vec{y}),
	\label{eq:cov_vec}
\end{equation}
where the components ${v^\alpha}_{;\beta}$ are given by 
\begin{equation}
	{v^\alpha}_{;\beta} := {v^\alpha}_{,\beta} + v^\gamma 
	\Gamma^\alpha_{\beta\gamma}.
	\label{eq:cov_components}
\end{equation}
Once again, a subscript comma ``,'' is shorthand for a partial derivative, 
${v^i}_{,j} := \partial v^i / \partial u^j$, whilst the 
$\Gamma^\alpha_{\beta\gamma} = g^{\alpha\delta}\left(g_{\delta\beta,\gamma} + 
g_{\delta\gamma,\beta} - g_{\beta\gamma,\delta}\right)/2$ are Christoffel 
symbols, which define the action of the covariant derivative, via 
$\nabla_{\vec{e}_\alpha} \vec{e}_\beta = \vec{e}_\gamma 
\Gamma^\gamma_{\alpha\beta}$.  Note that the shorthand $\nabla_\alpha := 
\nabla_{\vec{e}_\alpha}$ is frequently used in physics.

\subsubsection*{1-forms}
For a $1$-from, the action of the covariant derivative can be defined by 
demanding that the ``Leibniz rule'' holds.  That is, if a scalar field is 
defined by the action of a $1$-form on a vector, {\it i.e.}, 
$\phi:=v(\vec{w})=v^\alpha\,w_\alpha$, then
\begin{equation}
	\nabla_\alpha \left(v^\beta\,w_\beta\right) = 
	\left(v^\beta\,w_\beta\right)_\alpha := {v^\beta}_{;\alpha}\, w_\beta + 
	v^\beta \,w_{\beta;\alpha}.
	\label{eq:Leibnitz}
\end{equation}
The result is that $v_{\alpha;\beta}:= v_{\alpha,\beta} - 
v_\gamma\,\Gamma^\gamma_{\alpha\beta}$, 
which is consistent with the notion of using the metric as a raising / lowering 
operator ({\it i.e.}, $v_{\alpha;\beta}=g_{\alpha\gamma}{v^\gamma}_{;\beta}$).  
In coordinate free notation, this is equivalent to
\begin{equation}
	\left(\nabla_{\vec{y}} \,v \right) \left( \vec{w} \right) := 
	\nabla_{\vec{y}} \left[v\left( \vec{w} \right)\right] - v\left( 
	\nabla_{\vec{y}} \,\vec{w} \right).
	\label{eq:cov_1_form}
\end{equation}

\subsection*{Second fundamental form, Gauss and Weingarten 
equations}\label{sec:II}
Consider the derivative
\begin{equation}
	\frac{\partial\hat{\boldsymbol{n}}}{\partial 
	u^\alpha}=:\hat{\boldsymbol{n}}_{,\alpha},	\label{eq:dn_dualpha}
\end{equation}
{\it i.e.}, the rate (and direction) of change in the unit normal to 
$\mathcal{S}$ as $u^\alpha$ is varied, expressed as a vector in $\mathbb{R}^3$. Since $\hat{\boldsymbol{n}}$ is a unit vector, the result must still be tangent 
to $\mathcal{S}$ and therefore
\begin{equation}
	\hat{\boldsymbol{n}}_{,\alpha} = -{b^\beta}_\alpha 
	\,\boldsymbol{F}_\ast(\vec{e}_\beta),
	\label{eq:Wiengarten}
\end{equation}
which is known as the Weingarten equation (the assignment of a minus sign being 
convention). Given the right-hand side, we can use the coefficients from the 
above to construct a linear map $b:T\mathbb{R}^2\to T\mathbb{R}^2$ by writing 
$\vec{b}(\vec{v})=-v^\beta \vec{e}_\alpha\, {b^\alpha}_\beta$, for arbitrary 
$\vec{v}$.  Similarly, there is a natural bilinear form $\mathrm{II}$, know as 
the {\it second fundamental form}, that can be associated with such a map, 
whose action is given by
\begin{equation}
	\mathrm{II}\left(\vec{v},\vec{w}\right) = 
	\left\langle\vec{v},\vec{b}(\vec{w})\right\rangle =  v^\gamma 
	w^\beta\left\langle\vec{e}_\gamma,-{b^\alpha}_\beta\, \vec{e}_\alpha 
	\right\rangle.
	\label{eq:B}
\end{equation}
That is
\begin{equation}
	\mathrm{II} = - \left[\boldsymbol{F}_\ast 
	\left(\vec{e}_\alpha\right)\cdot\hat{\boldsymbol{n}}_{,\beta}\right]\,\mathrm{d}u^\alpha\otimes\mathrm{d}u^\beta 
	= b_{\alpha\beta}\,\mathrm{d}u^\alpha\otimes\mathrm{d}u^\beta,
	\label{eq:b_ij}
\end{equation}
where $b_{\alpha\beta}=g_{\alpha\gamma}b^\gamma_\beta = 
-\left[\boldsymbol{F}_\ast 
\left(\vec{e}_\alpha\right)\cdot\hat{\boldsymbol{n}}_{,\beta}\right]$.  Notice 
that since $\partial 
\left[\boldsymbol{F}_\ast(\vec{e}_\alpha)\cdot\hat{\boldsymbol{n}}\right] / 
\partial u^\beta = 0$, we have $b_{\alpha\beta} =  \left(\partial 
\boldsymbol{F}_\ast(\vec{e}_{\alpha})/\partial u^\beta\right)\cdot 
\hat{\boldsymbol{n}}$.  More generally, the derivative of basis vectors 
$\vec{e}_\alpha$ with respect to some coordinate $u^\beta$ on $\mathcal{S}$ can 
be decomposed into tangent and normal parts.
\begin{equation}
	\frac{\partial \boldsymbol{F}_\ast(\vec{e}_{\alpha})}{\partial u^\beta} = 
	\Gamma^{\gamma}_{\alpha\beta}\boldsymbol{F}_\ast(\vec{e}_\gamma) + 
	b_{\alpha\beta}\,\hat{\boldsymbol{n}},
	\label{eq:Gauss}
\end{equation}
which is known as Gauss' equation.

\subsection*{Curvature}\label{sec:curv}
\subsubsection*{Lines}
In $\mathbb{R}^3$, the curvature $q$ of a line $\boldsymbol{\gamma}(t)$ at a 
given point $x$ is just the norm of the covariant derivative acting on the unit 
tangent to the line (in the direction tangent to $\boldsymbol{\gamma}$) at $x$:
\begin{equation}
	q := \left\Vert
	\nabla_{\hat{\boldsymbol{T}}} \,\hat{\boldsymbol{T}} \right\Vert
	=\left[
		\delta_{ij} \left(
		\nabla_{\hat{\boldsymbol{T}}} \hat{\boldsymbol{T}}
		\right)^i
		\left(
		\nabla_{\hat{\boldsymbol{T}}} \hat{\boldsymbol{T}}
		\right)^j
	\right]^{1/2},
	\label{eq:q}
\end{equation}
where
\begin{equation}
	\hat{\boldsymbol{T}} = \left\Vert \frac{d \boldsymbol{\gamma}}{d
	t}\right\Vert^{-1} \frac{d \boldsymbol{\gamma}}{d t},
	\label{eq:T_def}
\end{equation}
are just normalised tangent vectors to $\boldsymbol{\gamma}(t)$ and the 
covariant derivative reduces to the usual directional derivative of 
$\mathbb{R}^3$, {\it i.e.}, $\hat{\boldsymbol{T}}\cdot\left( \partial/\partial 
x, \partial/\partial y, \partial/\partial z \right)^\mathsf{T}$.

For lines that are also embedded in a sub-manifold ({\it e.g.}, the boundary 
$\partial\mathcal{S}$ is a line in both $\mathbb{R}^3$ and $\mathcal{S}$) there 
are two common measures of curvature: geodesic and normal.  The geodesic 
curvature is a measure of curvature in the tangent plane $T\mathcal{S}$.  That 
is
\begin{equation}
	q_g := \left\Vert
	\nabla_{\hat{\vec{T}}} \,\hat{\vec{T}} \right\Vert
	=\left[
		g_{\alpha\beta} \left(
		\nabla_{\hat{\vec{T}}}\,\hat{\vec{T}}
		\right)^\alpha
		\left(
		\nabla_{\hat{\vec{T}}}\,\hat{\vec{T}}
		\right)^\beta
	\right]^{1/2},
	\label{eq:q_g_def}
\end{equation}
where $\hat{\vec{T}}$ is the pre-image of $\hat{\boldsymbol{T}}$ under 
$F_\ast$.  By contrast to (\ref{eq:q_g_def}), the normal curvature measures the 
out-of-(sub)manifold curvature and is given by
\begin{equation}
	q_n :=\left[
		\delta_{ij} \left\{\left[\left(
			\nabla_{\hat{\boldsymbol{T}}} \,\hat{\boldsymbol{T}}\right)
		\cdot\hat{\boldsymbol{n}}
	\right]\,\hat{\boldsymbol{n}}\right\}^i
		\left\{\left[
			\left(\nabla_{\hat{\boldsymbol{T}}} 
			\,\hat{\boldsymbol{T}}\right)\cdot
		\hat{\boldsymbol{n}}\right]\,\hat{\boldsymbol{n}}
	\right\}^j
	\right]^{1/2}.
	\label{eq:q_n}
\end{equation}
Both $q_g$ and $q_n$ are linked to $q$ via the following relation
\begin{equation}
	q = \sqrt{q_g^2 + q_n^2}.
	\label{eq:curv_link}
\end{equation}

\subsubsection*{Surfaces}
At a given point $p\in\mathbb{R}^2$, each unit vector $\hat{\vec{y}}$ 
corresponds to a unique curve $C$ on $\mathcal{S}$ that also lies in the plane 
$\mathcal{P}$ spanned by $\hat{\boldsymbol{n}}$ and 
$\boldsymbol{F}_\ast\left(\hat{\vec{y}}\right)$.  The action of the second 
fundamental form on a given $\hat{\vec{y}}$, results in the {\it normal} 
curvature $c^{(n)}_{\hat{\vec{y}}}$ of $\mathcal{S}$ in the direction of 
$\boldsymbol{F}_\ast\left(\hat{\vec{y}}\right)$ ({\it i.e.}, the curvature of 
$C$ in $\mathcal{P}$).  We write,
\begin{equation}
	\mathrm{II}(\hat{\vec{y}},\hat{\vec{y}}) = \pm c^{(n)}_{\hat{\vec{y}}},
	\label{eq:kappa}
\end{equation}
where ``+'' indicates whether $C$ is curving towards the unit normal, and {\it 
vice-versa} for ``-''.  Since the normal curvature will change dependent on 
which direction $\hat{\vec{y}}$ is chosen, we define the principal directions:
\begin{equation}
	\hat{\vec{y}}_1 (p) = \argmax_{\hat{\vec{y}}\in 
	T_p\mathbb{R}^2}\,\mathrm{II}(\hat{\vec{y}}, \hat{\vec{y}}),\ \mathrm{and}\ 
	\hat{\vec{y}}_2(p) = \argmin_{\hat{\vec{y}}\in T_p\mathbb{R}^2}\, 
	\mathrm{II}(\hat{\vec{y}}, \hat{\vec{y}}).
	\label{eq:T_12}
\end{equation}
The principal curvatures are then given by
\begin{equation}
	c_\alpha(p) = \mathrm{II}(\hat{\vec{y}}_\alpha,\hat{\vec{y}}_\alpha),\ 
	\forall\,  \alpha = 1,2.
	\label{eq:kappa_12}
\end{equation}
It can be shown that the $c_\alpha$ are eigenvalues of the linear operator 
$\vec{b}$ from \S \ref{sec:II}.  That is,
\begin{equation}
	\vec{b}(\hat{\vec{y}}_\alpha) = c_\alpha\,\hat{\vec{y}}_\alpha,
	\label{eq:princ}
\end{equation}
where if $c_1\neq c_2$, the principal directions are orthogonal.  We may now 
define two different types of local curvature of $\mathcal{S}$: the {\it mean} 
curvature 
\begin{equation}
	H := \frac{1}{2}\mathrm{Tr}\,{b^\alpha}_\beta = 
	\frac{1}{2}\mathrm{Tr}_g\,\mathrm{II} = \frac{c_1+c_2}{2},
	\label{eq:H}
\end{equation}
and the {\it Gaussian} curvature
\begin{equation}
	K:=\mathrm{Det}\,{b^\alpha}_\beta = 
	\frac{\mathrm{Det}\,b_{\alpha\beta}}{\mathrm{Det}\,g_{\alpha\beta}} = 
	c_1\,c_2.
	\label{eq:K}
\end{equation}
\section*{Solvated Membrane Nanodiscoids}\label{sec:prob_setup}
Consider a nanoscale bilayer discoid, stabilised by a solvating compound such 
as Styrene Maleic Acid or the lipo-protein APO-1A.  Write 
$\mathcal{H}=\mathcal{H}_m + \mathcal{H}_b$, where
\begin{equation}
	\mathcal{H}_m = \int_\mathcal{S} dA \left[ \sigma + \frac{\kappa}{2}\left( 
		2H - c_0
	\right)^2 + \bar{\kappa} K \right],
	\label{eq:H_m_SM}
\end{equation}
and
\begin{equation}
	\mathcal{H}_b = \int_{\partial\mathcal{S}} dl \left[ \tau + 
	\frac{k_g}{2}\left( q_g - q_0 \right)^2 + \frac{k_n}{2} q_n^2 \right],
	\label{eq:H_e_SM}
\end{equation}
are energetic contributions from the bilayer membrane and solvating compound, 
respectively.  In the first integral (\ref{eq:H_m_SM}), $dA$ is the surface area 
element of $\mathcal{S}$, $H$ and $K$ are the mean [{\it cf.}~Eq.~(\ref{eq:H})] 
and Gaussian [{\it cf.}~Eq.~(\ref{eq:K})] curvatures respectively, and $\sigma$ 
is a surface tension.  The moduli $\kappa$ and $\bar{\kappa}$ are the bending 
and Gaussian (saddle-splay) rigidities, respectively, and $c_0$ is the 
membrane's spontaneous (mean) curvature.  In the second integral 
(\ref{eq:H_e_SM}), $dl$ is the line element [{\it cf.}~Eq.~(\ref{eq:dl_def})] of 
the surface boundary $\partial\mathcal{S}$, $\tau$ is a line tension, $q_g$ is 
the geodesic curvature [{\it cf.}~Eq.~(\ref{eq:q_g_def})], and $q_n$ is the 
normal curvature [{\it cf.}~Eq.~(\ref{eq:q_n})].  The coefficients $k_g$ and 
$k_n$ are corresponding bending moduli with dimensions of energy multiplied by 
length.  The SMA co-polymer is assumed to have spontaneous (line) curvature 
$q_0$ in the tangent plane to the disc, but not in the normal direction 
(\textit{i.e.}, it preserves the up-down symmetry of the bilayer).  Restricting 
the analysis to surfaces with an Euler characteristic of one, the Gauss-Bonnet 
theorem gives:
\begin{equation}
	\int_\mathcal{S} K\,dA = \int_{\partial\mathcal{S}} k_g\,dl + 
	2\pi.
	\label{eq:G-B}
\end{equation}
Substituting into (\ref{eq:H_m_SM}) and (\ref{eq:H_e_SM}), we have
\begin{equation}
	\mathcal{H} = \int_\mathcal{S} dA \left[ \sigma + \frac{\kappa}{2}\left( 2H 
		- c_0
	\right)^2 + \bar{\kappa}_g K \right] + \int_{\partial\mathcal{S}} dl \left[ 
	\tau_g + \frac{k_g}{2}\, q_g^2 + \frac{k_n}{2}\,q_n^2 \right] - 
	k_g\,q_0\,2\pi.
\label{eq:H_RN}
\end{equation}
where $\bar{\kappa}_k:=\bar{\kappa} + q_0\,k$ and $\tau_k := \tau + k\,q_0^2 / 
2$ are re-normalised constants.

\subsection*{Flat}\label{sec:flat}
Assume that $H=K=c_0=q_n=0$, which implies $\mathcal{S}$ is just some bounded 
domain in $\mathbb{R}^2$ and only the line integral of (\ref{eq:H_RN}) needs to 
be calculated.  Using polar coordinates $(r,\theta)$ it is assumed that the 
discoid boundary $\partial\mathcal{S}$ can be characterised by the vector field 
$\vec{\gamma}(\theta)$, which is taken to be single-valued.  Throughout, we use 
the convention that the magnitude of a vector is indicated by omitting the 
overarrow, \textit{i.e.}, $\gamma(\theta)=\Vert \vec{\gamma}(\theta)\Vert$.  In 
the natural basis $\vec{e}_1 = \hat{\vec{r}}$ and $\vec{e}_2 = 
r\hat{\vec{\theta}}$, the set of vectors tangent to the line 
$\vec{\gamma}(\theta)$, are just given by $d\vec{\gamma}/d\theta = 
\gamma\vec{e}_1 + \vec{e}_2$.  The metric is just that of $\mathbb{R}^2$ in 
polar coordinates, {\it i.e.}, $g_{\alpha\beta} = \mathrm{Diag}[1,r^2]$.  The 
``line element'' $dl$ [{\it cf.}~Eq.~(\ref{eq:dl_def})] is then given by
\begin{equation}
		dl=\left[ \gamma^2 + \left( \frac{d\gamma}{d\theta} 
		\right)^2\right]^{1/2} d\theta.
	\label{eq:dl_2}
\end{equation}
Similarly, the normalised vectors tangent to $\partial\mathcal{S}$ are given by
\begin{equation}
	\hat{\vec{T}} =\frac{1}{\sqrt{\gamma^2 + \left( d\gamma/d\theta 
	\right)^2}}\left( \frac{d \gamma}{d\theta},1\right)^\mathsf{T}.  
	\label{eq:T}
\end{equation}
Here, since the Christoffel symbols ${\Gamma}{^i_{jk}}$ are only non-zero in 
three cases: ${\Gamma}{^1_{22}} = -r$, and ${\Gamma}{^2_{12}} = 
{\Gamma}{^2_{21}} = 1/r$, the components of the covariant derivative 
${\nabla}{_{\hat{\vec{T}}}}\,\hat{\vec{T}}$ are
\begin{equation}
	\left( {\nabla}{_{\hat{\vec{T}}}}\,\hat{\vec{T}} \right)^1=
	-\frac{\gamma\, \left[ \gamma^2 + 2\left(\gamma^\prime \right)^2 - \gamma\, 
	\gamma^{\prime\prime}  \right]}{\left[ \gamma^2 + \left( \gamma^\prime 
	\right)^2 \right]^{2}},\quad \mathrm{and}\quad
	\left( {\nabla}{_{\hat{\vec{T}}}}\,\hat{\vec{T}} \right)^2=
				\frac{ \gamma^\prime\,\left[ \gamma^2 + 2\left(\gamma^\prime 
				\right)^2 - \gamma\, \gamma^{\prime\prime}  
			\right]}{\gamma\,\left[ \gamma^2 + \left( \gamma^\prime \right)^2 
			\right]^{2} },
	\label{eq:DT_Ds^}
\end{equation}
where the shorthand $\gamma'=d\gamma/d\theta$ has been introduced for 
readability.  Returning to (\ref{eq:q_g_def}), we see that
\begin{equation}
	q_g = \frac{\gamma^2 + 2\left(\gamma^\prime \right)^2 - 
		\gamma\,\gamma^{\prime\prime}}{\left[ \gamma^2 + \left( \gamma^\prime 
		\right)^2 \right]^{3/2}}.
	\label{eq:q_r_2}
\end{equation}
If the membrane boundary is quasi-circular, then $\gamma(\theta) = R_0 \left[1 
+ \epsilon f(\theta) \right]$, where $f(\theta)=\sum_n \Re\left[A_n \exp\left( 
i\,n\, \theta\right) \right]$, such that $\Re$ is the real part, $A_n \in 
\mathbb{C},\ \vert A_n\vert \le 1\ \forall\ n\in\mathbb{N}$, and $\epsilon \ll 
1$.  A power series expansion in $\epsilon$ can then be performed on 
Eqs.~(\ref{eq:dl_2}) and (\ref{eq:q_r_2}), with the results
\begin{equation}
	\begin{split}
		dl = R_0
		\left\{
			1 + \epsilon\,\sum_n^\infty\Re
			\left[
				A_n \exp\left(i\,n\,\theta\right)
			\right]
			+ \frac{\epsilon^2}{2}\sum_n^\infty n\,\Re
			\left[
				i\,A_n\,\exp\left(i\,n\,\theta\right)
			\right]^2
		\right\}\,d\theta + O\left( \epsilon^3 \right),
		\label{eq:dl_pert}
	\end{split}
\end{equation}
and
\begin{equation}
	\begin{split}
		q_g=\frac{1}{R_0}
		\left\{
			1 + \epsilon\,\sum_n^\infty\left( n^2-1 \right)\,\Re
			\left[
				\exp\left( i\,n\,\theta \right)
			\right]
			-\frac{\epsilon}{4}\sum_n^\infty
			\left\{
				\left\vert A_n \right\vert^2
				\left( 3n^2 - 2 \right)
				+
				\Re
				\left[
					A_n^2\exp\left( 2\,i\,n\,\theta \right)
				\right]
			\right\}
		\right\} + O\left(\epsilon^3\right).
	\label{eq:qr_pert}
	\end{split}
\end{equation}
The energy $\mathcal{H}$ can then be shown to be of the form
\begin{equation}
	\begin{split}
	\mathcal{H} =& \,\pi\left[ \frac{k_g\left( q_0 R_0 - 1 \right)^2}{R_0}
		+ R_0\left( R_0 \sigma + 2\tau \right) \right]\\
		&+ \epsilon^2\frac{\pi}{4 R_0}\Big\{ k_g\left[ 2 + \left( q_0^2\, R_0^2 
		- 5 \right) n^2 + 2n^4\right] + 2 R_0^2\left( R_0\,\sigma + n^2\tau 
	\right)\Big\}\sum_n \left\vert A_n\right\vert^2 + O\left( \epsilon^3 
	\right).
\end{split}
	\label{eq:H_disc_pert}
\end{equation}
The minimum of $\mathcal{H}$ is given by requiring $\partial \mathcal{H} / 
\partial R_0 \vert_{\epsilon=0} = 0$.  The result is that
\begin{equation}
	\tau = \frac{k_g}{2R_0^2}\left( 1 - q_0^2\,R_0^2\right) - R_0\,\sigma,
	\label{eq:tau_SM}
\end{equation}
which implies
\begin{equation}
	\mathcal{H} = \pi\left[\frac{k_g}{R_0}\left( 1 - q_0\,R_0 \right) - 
	\frac{R_0^2\,\sigma}{2}\right] + \frac{\pi\,\epsilon^2}{2\,R_0}\sum_n 
	\left( n^2 - 1 \right) \left[ k_g \left( n^2 - 1 \right) - R_0^3\,\sigma 
	\right] \left\vert A_n\right\vert^2 + O\left( \epsilon^3 \right).
\label{eq:H^_min}
\end{equation}
Notice that terms of $O\left( \epsilon^2 \right)$ in Eq.~(\ref{eq:H^_min}) do 
\emph{not} rely on $q_0$, and therefore neither does the mean-squared amplitude 
of each mode in equilibrium.

The principle of equipartition of energy states that each quadratic mode 
contributes $ k_\mathrm{B} T / 2$ to the expectation value of the energy--- 
{\it i.e.}, summing $\mathcal{H}$ over all configurations $\{A_n: n\in 
	\mathbb{N}\}$, weighted by the Boltzmann distribution. Using angle brackets 
	to indicate expectation value, this implies $\left\langle 
	\mathcal{H}_n\right\rangle = k_\mathrm{B} T / 2$, where
\begin{equation}
	\mathcal{H}_n = \frac{\pi\,\epsilon^2}{2\,R_0}\left( n^2 - 1 \right) \left[ 
	k_g \left( n^2 - 1 \right) - R_0^3\,\sigma \right]\left\vert 
	A_n\right\vert^2.
\end{equation}
Equation (3) in the main text then follows from the above.

\subsection*{Non-flat}\label{sec:nonflat}
Parameterise the membrane shape by using the two (orthogonal) principal 
curvatures, chosen without loss of generality such that $\vert c_1\vert \ge 
\vert c_2\vert$.  The height field in an polar Monge approach is given by
\begin{equation}
		h(r,\theta) = \frac{r^2}{4}\left[ c_1 + c_2 + \left( c_1 - c_2 \right) 
		\cos 2\theta \right],
	\label{eq:h}
\end{equation}
where the angle $\theta$ is assumed to be measured from the principal axis 
associated with $c_1$.  Moreover, $\left\Vert \nabla h \right\Vert \ll 1\ 
\forall\ (r,\theta)$, therefore
\begin{equation}
	\max\left\{\left\Vert \nabla h \right\Vert:\theta\in[0,2\pi), 
	r\in[0,\gamma(\theta)]\right\} = \left\Vert \nabla h 
	\right\Vert_{\theta=0,\,r=\gamma(0)} = \gamma(0)\,c_1 \ll 1
	\label{eq:monge}
\end{equation}
Since $\gamma(\theta)=R_0\left[ 1 + \epsilon f(\theta) \right]$, we formally 
set $\alpha = R_0\,c_1 \ll 1$, therefore (\ref{eq:h}) becomes
\begin{equation}
	h(r,\theta) = \alpha\,\psi(r,\theta),\ \mathrm{with} \ \psi(r,\theta) = 
	\frac{r^2}{4\,R_0}\left[ 1 + \phi + \left(1-\phi \right) \cos 2 \theta
	\right],
	\label{eq:h_alpha_SM}
\end{equation}
where $\phi=c_2/c_1$ takes values in the interval $[-1,1]$.  Making contact 
with \S\ref{sec:setup}, $\mathcal{S}$ is parameterised by 
$\{u^\alpha:\alpha=1,2\}$ and $\mathbb{R}^3$ by $\{x^i:1=1,2,3\}$, where a 
Monge gauge is tantamount to choosing the map $\boldsymbol{F}(u) = (u^1, u^2, 
h(u^1,u^2))^\mathsf{T}$.  Moreover, $u^1 = r$ and $u^2 = \theta$, which implies 
that the embedding basis $\{\boldsymbol{e}_i:i=1,2,3\}$ is just that of 
cylindrical polars 
$\{\hat{\boldsymbol{r}},r\hat{\boldsymbol{\theta}},\hat{\boldsymbol{z}}\}$.  
The position vector is then $\boldsymbol{R}(r,\theta) := r \hat{\boldsymbol{r}} 
+ \alpha\,\psi(r,\theta)\hat{\boldsymbol{z}}$, and tangent vectors to the 
surface are spanned by the set:
\begin{equation}
\boldsymbol{F}_\ast\left(\vec{e}_1\right) = 
\frac{\partial\boldsymbol{R}}{\partial r} = \hat{\boldsymbol{r}} + 
\alpha\,\frac{\partial \psi}{\partial r} \hat{\boldsymbol{z}},\ \mathrm{and}\ 
\boldsymbol{F}_\ast\left(\vec{e}_2\right) = 
\frac{\partial\boldsymbol{R}}{\partial \theta} = r\hat{\boldsymbol{\theta}} + 
\alpha\,\frac{\partial \psi}{\partial \theta} \hat{\boldsymbol{z}}.
	\label{eq:e_i_monge}
\end{equation}
Recalling the shorthand $\psi_{,\alpha} := \partial \psi / \partial u^\alpha$, 
the metric and its inverse become
\begin{equation}
	{g}{_{\alpha\beta}} = \left\langle
	\vec{e}_\alpha,\vec{e}_\beta \right\rangle = \boldsymbol{F}_\ast \left( 
	\vec{e}_\alpha \right)\cdot\boldsymbol{F}_\ast\left( \vec{e}_\beta \right) 
	= \left(
		\begin{array}{cc}
			1 + \alpha^2\,\left( \psi_{,1} \right)^2 & 
			\alpha^2\,\psi_{,1}\,\psi_{,2}\\
			\alpha^2\,\psi_{,1}\,\psi_{,2} & r^2 + \alpha^2\,\left( \psi_{,2} 
			\right)^2\end{array}
		\right),
		\label{eq:tilde_g_ij_polars}
\end{equation}
and
\begin{equation}
	{g}{^{\alpha\beta}} =\left({g}{_{\alpha\beta}}\right)^{-1}
	=\frac{1}
	{r^2\left( 1 +
		\alpha^2 \left\Vert \nabla^{\mathbb{R}^2}\psi\right\Vert^2 \right)}
	\left(
	\begin{array}{cc}
		r^2 + \alpha^2\,\left( \psi_{,2} \right)^2 & 
		-\alpha^2\,\psi_{,1}\,\psi_{,2} \\
		-\alpha^2\,\psi_{,1}\,\psi_{,2} & 1 + \alpha^2\,\left( \psi_{,1} 
		\right)^2\end{array}
	\right).
	\label{eq:tilde_g^ij_polars}
\end{equation}
It is clear that ${g}{ } := \mathrm{Det}\left({g}{_{\alpha\beta}}\right) = 
r^2\left( 1 + \left\Vert \nabla^{\mathbb{R}^2} \psi\right\Vert^2 \right)$, 
where $\nabla^{\mathbb{R}^2}$ is the covariant derivative in $\mathbb{R}^2$ 
(polar coordinates are assumed).  We may also calculate the normal to the 
surface [{\it cf.}~Eq.~(\ref{eq:nhat})],
\begin{equation}
	\boldsymbol{n} = \hat{\boldsymbol{z}} \left( 1 - \alpha^2
	\left\Vert \nabla^{\mathbb{R}^2} \psi\right\Vert^2 \right) - 
	\hat{\boldsymbol{r}}\,\alpha\, \psi_{,1} - 
	\hat{\boldsymbol{\theta}}\,\alpha\,\psi_{,2} + O\left( \alpha^3 \right),
	\label{eq:n_pert}
\end{equation}
and the derivatives of the tangent vectors (pushed forwards to $\mathbb{R}^3$), 
{\it i.e.},
\begin{equation}
	\frac{\partial \boldsymbol{F}_\ast\left( \vec{e}_1 \right)}{\partial u^1} = 
	\frac{\partial^2\boldsymbol{R}}{\partial r^2} = 
	\alpha\,\psi_{,11}\,\hat{\boldsymbol{z}},\ \frac{\partial 
		\boldsymbol{F}_\ast\left( \vec{e}_2 \right)}{\partial u^2}= 
		\frac{\partial^2\boldsymbol{R}}{\partial \theta^2} = 
		-r\,\hat{\boldsymbol{r}} + \alpha\,\psi_{,22}\,\hat{\boldsymbol{z}},\ 
		\mathrm{and}\ \frac{\partial \boldsymbol{F}_\ast\left( \vec{e}_1 
	\right)}{\partial u^2} = \frac{\partial^2 \boldsymbol{R}}{\partial 
	r\partial \theta} = \hat{\boldsymbol{\theta}} + 
	\alpha\,\psi_{,12}\,\hat{\boldsymbol{z}},
	\label{eq:second}
\end{equation}
from which the coefficients of the second fundamental form (\ref{eq:b_ij}) may 
be constructed:
\begin{equation}
	b_{\alpha\beta} =
	\alpha\,\left(
	\begin{array}{cc}
		\psi_{,11} & \psi_{,12} - \psi_{,2}/r \\
		\psi_{,12} - \psi_{,2}/r &  r\,\psi_{,1} + \psi_{,22}
	\end{array}
	\right) + O\left( \alpha^3 \right).
	\label{eq:L_ij}
\end{equation}
Using the definitions (\ref{eq:H}) and (\ref{eq:K}) it can be shown that, up to 
second order in the small parameter $\alpha$, the mean and Gaussian curvatures 
are just the trace and determinant of the (polar coordinate) Hessian of $\psi$.  
That is
\begin{equation}
	H = \mathrm{Tr}\left[ 
	\alpha\,\mathrm{Hess}^{\mathbb{R}^2}\left(\psi\right)\right]+ O\left( 
	\alpha^3 \right) = \frac{\alpha}{2}\left( \psi_{,11} + \frac{\psi_{,1}}{r} 
	+ \frac{\psi_{,22}}{r^2} \right)+ O\left( \alpha^3 \right) = 
	\frac{\alpha}{2}\,^{\mathbb{R}^2}\nabla^2 \psi+ O\left( \alpha^3 \right),
	\label{eq:H_psi}
\end{equation}
and
\begin{equation}
	K = \frac{1}{r^2}\mathrm{Det}\left[ 
	\alpha\,\mathrm{Hess}^{\mathbb{R}^2}\left(\psi \right)\right]+ O\left( 
	\alpha^3 \right) = \frac{\alpha^2}{r^2}\left[ \psi_{,11}\left( r\,\psi_{,1} 
	+ \psi_{,22} \right)-\left( \psi_{,12} - \frac{\psi_{,2}}{r} 
\right)^2\right]+ O\left( \alpha^3 \right),
	\label{eq:K_psi}
\end{equation}
where
\begin{equation}
	\left[ \mathrm{Hess}^{\mathbb{R}^2}\left( \psi \right) 
	\right]_{\alpha\beta} :=
	\psi_{,\alpha;\beta} = \psi_{,\alpha\beta} - 
	^{\mathbb{R}^2}\Gamma^\gamma_{\alpha\beta}\,\psi_{,\gamma}.
	\label{eq:Hess}
\end{equation}
Substituting for (\ref{eq:h_alpha_SM}) leads to the results:
\begin{equation}
	H = \frac{\alpha}{R_0}\left( 1 + \phi \right) + O\left( \alpha^3 \right),\ 
	\mathrm{and}\ K = \frac{\phi\,\alpha^2}{R_0^2} + O\left( \alpha^3 \right).
	\label{eq:HandK_subst}
\end{equation}
That is, up to $O\left( \alpha^2 \right)$, the mean and Gaussian curvatures of 
a perturbation of the form (\ref{eq:h_alpha_SM}) are constant.

In a similar way to the above treatment of $H$ and $K$, both the line element 
$dl$, and curvatures $q_g$ and $q_n$ may be expanded in terms of $\alpha$.  
However, these quantities also rely on $\epsilon$.  To avoid confusion between 
power series expansions of $\alpha$ and $\epsilon$, we adopt the notation that 
coefficients are labelled in the following way: in an expansion of some 
function $F$, the $O\left( \alpha^a\,\epsilon^e \right)$ term is written as
$\alpha^a\,\epsilon^e\,F^{(a,e)}$.  For example, using this convention, the 
results of the previous section can be re-labelled.  The terms of 
(\ref{eq:dl_pert}) become
\begin{equation}
	dl^{(0,0)} = R_0\,d\theta,\ dl^{(0,1)} = R_0\,\sum_n^\infty\Re
			\left[
				A_n \exp\left(i\,n\,\theta\right)\,d\theta
			\right],\ \mathrm{and}\ dl^{(0,2)} = \frac{R_0}{2}\sum_n^\infty 
			n\,\Re
			\left[
				i\,A_n\,\exp\left(i\,n\,\theta\right)
			\right]^2\,d\theta,
	\label{eq:dl_pert_2}
\end{equation}
while the terms of (\ref{eq:qr_pert}) are given by
\begin{equation}
	q_g^{(0,0)} = \frac{1}{R_0},\ q_g^{(0,1)} = \frac{1}{R_0} 
	\sum_n^\infty\left( n^2-1 \right)\,\Re
			\left[
				\exp\left( i\,n\,\theta \right)
			\right],
			\label{eq:qr_pert_2_O_0_1}
\end{equation}
and
\begin{equation}
	q_g^{(0,2)} = \frac{-1}{4\,R_0}\sum_n^\infty
	\Big\{
		\left\vert A_n \right\vert^2
		\left( 3n^2 - 2 \right)
		+
		\Re
		\left[
			A_n^2\exp\left( 2\,i\,n\,\theta \right)
		\right]
	\Big\}.
\label{eq:qr_pert_2_O2}
\end{equation}
The full expansions, now in terms of both $\alpha$ and $\epsilon$, take the 
form
\begin{equation}
	dl=dl^{(0,0)} + \epsilon\,dl^{(0,1)} + \epsilon^2\,dl^{(0,2)} + 
	\alpha^2\,dl^{(2,0)} + O\left(\epsilon^3\right) + O\left( \alpha^3 \right),
	\label{eq:dl_pert_both}
\end{equation}
and
\begin{equation}
	q_g=q_g^{(0,0)} + \epsilon\,q_g^{(0,1)} + \epsilon^2\,q_g^{(0,2)} + 
	\alpha^2\,q_g^{(2,0)} + O\left(\epsilon^3\right) + O\left( \alpha^3 
	\right),
	\label{eq:qr_pert_both}
\end{equation}
where
\begin{equation}
	dl^{(1,0)} = 0,\ \mathrm{and}\ dl^{(2,0)} = 
	\frac{\psi_{,2}(R_0,\theta)^2}{2\,R_0}=\frac{R_0}{8}\left( \phi-1 \right)^2 
	\sin^2 \left( 2\theta \right),
	\label{eq:dl_pert_alpha}
\end{equation}
and
\begin{equation}
	q_g^{(1,0)} = 0,\ \mathrm{and}\ q_g^{(2,0)} =\frac{1}{2\,R_0^3}\left[ 
	-2\,\psi_{,2}(R_0,\theta)^2 + \psi_{,22}(R_0,\theta)^2 \right]= 
	\frac{1}{8\,R_0^2}\left(1-\phi  \right)^2 \left[ 1 + 3\cos\left(4\theta 
		\right)
\right].
	\label{eq:qr_pert_alpha}
\end{equation}
For the normal curvature, we have $q_n = \alpha\,q_n^{(1,0)} + O\left( \alpha^3 
\right) + O\left(\epsilon^3\right)$, where
\begin{equation}
	q_n^{(1,0)} = \frac{1}{R_0^2}\left[ \psi_{,22}(R_0,\theta) + 
	R_0\,\psi_{,1}(R_0,\theta) \right] = \frac{1}{2\,R_0}\left[ 1 + \phi + 
	\left( \phi - 1 \right)\cos\left( 2\,\theta \right) \right].
	\label{eq:q_n^10}
\end{equation}
Substituting the above results into Eq.~(\ref{eq:H_RN}), the necessary 
integrals can be performed in order to obtain an expansion of $\mathcal{H}$ in 
terms of both $\alpha$ and $\epsilon$.  (Note: the manipulations are quite 
tedious and we used the commercial software {\it Mathematica} \cite{Mma}).  If 
the spontaneous (mean) curvature $c_0$ is nonzero, then the expansion contains 
a term
\begin{equation}
	\mathcal{H}^{\left( 1,0 \right)} = -2\,c_0\,\kappa\,\pi\,R_0\left( 1 + \phi 
	\right),
	\label{eq:nonzero_c0}
\end{equation}
In this case, the up / down symmetry of the problem is broken, and {\it all} 
perturbations of the form (\ref{eq:h_alpha_SM}), other than a symmetric saddle, 
are unstable. (A symmetric saddle is given by $c_1 = -c_2$, such that $\phi = 
-1$ and $H=0$).  By contrast, if the spontaneous curvature is zero, then the 
energy is of the form
\begin{equation}
	\mathcal{H} = \mathcal{H}^{(0,0)} + \epsilon^2\, \mathcal{H}^{(0,2)} + 
	\alpha^2\,\mathcal{H}^{(2,0)} + O\left(\epsilon^3\right) + O\left( \alpha^3 
	\right).
	\label{eq:H_general_SM}
\end{equation}
Here, the line tension $\tau$ can once again be fixed for a given $R_0$ (and 
material parameters $k_g$, $k_n$, $q_0$ and $\sigma$) by imposing $\partial 
\mathcal{H}/\partial R_0\vert_{\epsilon=\alpha=0} = 0$.  The result is 
unchanged from (\ref{eq:tau_SM}).  In this case $\mathcal{H}^{(0,0)}$ and 
$\mathcal{H}^{(0,2)}$ can then be read-off from Eq.~(\ref{eq:H^_min}), {\it 
i.e.},
\begin{equation}
	\mathcal{H}^{(0,0)} = \pi\left[\frac{k_g}{R_0}\left( 1 - q_0\,R_0 \right) - 
	\frac{R_0^2\,\sigma}{2}\right], \mathrm{and}\ \mathcal{H}^{(0,2)} 
	=\frac{\pi}{2\,R_0}\left( n^2 - 1 \right) \left[ k_g \left( n^2 - 1 \right) 
	- R_0^3\,\sigma \right]\sum_n \left\vert A_n\right\vert^2.  \label{eq:H^0}
\end{equation}
The coefficient of $\alpha^2$ in (\ref{eq:H_general_SM}) is given by
\begin{equation}
	\mathcal{H}^{(2,0)} = \pi\,\left\{ \left( 1 + \phi \right)^2 2\,\kappa + 
		\left( 1-\phi \right)^2\frac{3\,k_g}{8\,R_0} +  \left[\left( 1+\phi 
		\right)^2 + 2\,\left( 1 + \phi^2 \right) \right]\,\frac{k_n}{8\,R_0} + 
		\phi\,\left( \frac{R_0^2\,\sigma}{4} + \bar{\kappa} + q_0\,k_g \right) 
	\right\},
	\label{eq:H^20}
\end{equation}
which can be re-written in the form $\mathcal{H}^{(2,0)}=\pi\,\phi\,\left[ 
\bar{\kappa} - \bar{\kappa}^\ast (\phi) \right]$, where 
\begin{equation}
	\begin{split}
		\bar{\kappa}^\ast &=
		-\frac{1}{\phi}\left( 1 + \phi \right)^2 2\,\kappa
		- \frac{1}{\phi}\left( 1 - \phi \right)^2 \frac{3\,k_g}{8\,R_0}
		- \frac{R_0^2\,\sigma}{4}\\
		&\quad- \frac{1}{\phi}
		\left[
			\left(1+\phi \right)^2 + 2\left( 1 +\phi^2\right)
		\right]\frac{k_n}{8\,R_0}
		 - q_0\,k_g,
	\end{split}
	\label{eq:kappabar_c}
\end{equation}
determines the stability of out-of-plane perturbations.  Setting $\phi = -1$ 
recovers
\begin{equation}
	\bar{\kappa}^\ast_p = \bar{\kappa}^\ast(-1) = \left(\frac{3}{2\,R_0} 
	-q_0\right)k_g + \frac{k_n}{2\,R_0} -\frac{R_0^2\,\sigma}{4},
	\label{eq:kappabar_p_SM}
\end{equation}
as shown in the main text.

\subsubsection*{Higher order contributions}
Consider the effect of contributions to the energy of order greater than 
$\alpha^2$.  In the expansion of $\mathcal{H}$, the terms of next lowest order 
can be calculated, and are at order $\alpha^4$.  However, to this order, 
$\epsilon$ and $\alpha$ do not de-couple, and we must explicitly set $\epsilon 
= 0$.  [In doing so, we revert to the simple notation that $O(\alpha^n)$ terms 
	in a series expansion of a given function, say $F$, are written 
$\alpha^n\,F^{(n)}$].  In addition, when expanding to $O\left( \alpha^4 
\right)$, the Helfrich Hamiltonian must be modified since Eq.~(\ref{eq:H_m_SM}) 
retains only lowest order terms ($\alpha^2$) by construction.  In our 
framework--- {\it i.e.}, shapes described by the polar monge field 
(\ref{eq:h_alpha_SM})--- the requisite higher order terms can be computed, and are 
given by
\begin{align}
	H^4 &= \frac{\alpha^4}{16\,R_0^4}\left( 1 + \phi \right)^4 + 
	O\left(\alpha^5\right)\label{eq:O41}\\
	H^2\,K &= \frac{\alpha^4}{4\,R_0^4}\,\phi\left( 1 + \phi \right)^2+ 
	O\left(\alpha^5\right),\\
	K^2 &= \frac{\alpha^4}{R_0^4}\,\phi^2+ O\left(\alpha^5\right),\\
	\Delta H^2 &= -\frac{\alpha^4}{2\,R_0^4} \left( 1+\phi \right)^2\left[ 
		3\left(1 + \phi^2
		\right) - 2\phi \right]+ O\left(\alpha^5\right),\\
	\Delta K &= -\frac{4\,\alpha^4}{R_0^4} \phi \left(1 + \phi^2\right)+ 
	O\left(\alpha^5\right).\label{eq:O45}
\end{align}
When multiplied by their respective moduli and integrated, these terms appear 
alongside other $O(\alpha^4)$ terms, which arise from the expansions of 
(\ref{eq:H_m_SM}) and (\ref{eq:H_e_SM}).  Assigning the coefficients $\lambda_i$ (for 
$i=1\dots5$) to the contributions (\ref{eq:O41}) to (\ref{eq:O45}), 
respectively, the resultant fourth order contribution to the energy 
$\mathcal{H}^{(4,0)}$ is given by
\begin{equation}
	\mathcal{H}^{(4,0)} = -\frac{3\pi}{8}\phi\left( 1+ \phi^2 
	\right)\left[\bar{\kappa}-\bar{\kappa}_p^{\dagger} \right],
	\label{eq:4th}
\end{equation}
where
\begin{equation}
	\begin{split}
	\bar{\kappa}^{\dagger} = \frac{8}{3\,\phi\left( 1+\phi^2 \right)}
	\Bigg\{&
		\kappa\left[ -\frac{5}{4} - \frac{3\phi}{2} - 
		\frac{\phi^2}{2}-\frac{2\phi^3}{2} - \frac{5\phi^4}{4}\right]		+
		\frac{k_n}{R_0}\left[ - \frac{31}{256} - 
		\frac{5\phi}{64}-\frac{77\phi^2}{128}-\frac{5\phi^3}{64} 
	-\frac{31\phi^4}{256}\right]
		\\&
		+
		\frac{k_g}{R_0}\left[ -\frac{19}{512} - \left(\frac{3 q_0 R_0}{8}- 
			\frac{19}{128}\right)\phi - \frac{57\phi^2}{256} - \left(\frac{3 
			q_0 R_0 }{8} - \frac{19}{128}\right)\phi^3 - \frac{19\phi^4}{512}
		\right]
		\\&
		+
		\sigma\,R_0^2\left[ -\frac{5}{512} - \frac{3\phi}{128} + 
			\frac{19\phi^2}{768} - \frac{3\phi^3}{128} - \frac{5\phi^4}{512} 
		\right] + \lambda_3\,\frac{\phi^2}{R^2_0} - 
		\lambda_5\,\frac{4\,\phi\left(1+\phi^2 \right)}{R_0^2}\\&
		+
		\frac{\left(1+\phi \right)^2}{R_0^2}\left[ \lambda_1\,\frac{\left( 
		1+\phi \right)^2}{16} + \lambda_2\,\frac{\phi}{4} - 
		\lambda_4\,\left(\frac{3}{2}\left( 1+\phi^2 \right) - \phi 
	\right)\right]
	\Bigg\}.
\end{split}
	\label{eq:kappa_p^dagger}
\end{equation}
Focusing, as before, on saddles with principal curvatures of equal magnitude 
(so-called ``pringles'') we set $\phi = -1$, resulting in 
\begin{equation}
	\bar{\kappa}_p^\dagger = \frac{19}{24}\frac{k_g}{R_0} + 
	\frac{11}{12}\frac{k_n}{R_0} - k_g q_0 - \frac{5}{72} R_0^2 \sigma - 
	\frac{4}{3}\frac{\lambda_3}{R_0^2} - \frac{32}{3}\frac{\lambda_5}{R_0^2},	
	\label{eq:kappaber_p^dag}
\end{equation}
where only $\Delta K$ and $K^2$ ({\it i.e.}, the terms involving Gaussian 
curvature) contribute from the higher order modifications (\ref{eq:O41}) to 
(\ref{eq:O45}). The coefficients in the expansion of the energy are now given 
by $\mathcal{H}^{(2)} = -\pi\,\left( \bar{\kappa} - \bar{\kappa}^\ast_p \right)$ 
and $\mathcal{H}^{(4)} = 3\,\pi\,\left( \bar{\kappa} - \bar{\kappa}^\dagger_p 
\right)/4$.  Due to the introduction of order $\alpha^4$ contributions, there 
are now solutions to $\partial\mathcal{H}/\partial \alpha = 0$ at non-zero 
$\alpha$, given by
\begin{equation}
	\alpha^\ast = \pm \left( \frac{-\mathcal{H}^{(2)}}{2\,\mathcal{H}^{(4)}} 
	\right)^{-1/2}.
	\label{eq:alphastar}
\end{equation}
The stability is determined by the coefficient $\mathcal{H}^{(2)}$, via
\begin{equation}
	\left.\frac{\partial^2 \mathcal{H}}{\partial 
	\alpha^2}\right\vert_{\alpha=\alpha^\ast}=-4\,\mathcal{H}^{(2)},
	\label{eq:alphastar_stab}
\end{equation}
which implies that $\alpha^\ast$ is only stable if 
$\bar{\kappa}>\bar{\kappa}^\ast_p$.  In addition, from (\ref{eq:alphastar}), 
the expression $-\mathcal{H}^{(2)} / 2\,\mathcal{H}^{(4)}$ must be positive, 
and we can therefore deduce that 
$\bar{\kappa}>\mathrm{max}\left(\bar{\kappa}^\ast_p,\bar{\kappa}^\dagger_p\right)$ 
for stable pringles.  

\subsubsection*{Sorting by modulus of Gaussian Rigidity}
Consider a two component nanodiscoid, where the each component has a different 
modulus of Gaussian rigidity, $\bar{\kappa}_1$ and $\bar{\kappa}_2$.  If $\mu$ 
is an area fraction, we write $\bar{\kappa} = \mu\, \bar{\kappa}_1 + (1-\mu)\, 
\bar{\kappa}_2$ as the average modulus of Gaussian rigidity of the (well mixed) 
membrane from which the discoids are cut.  Relative to the bulk membrane, a 
pringle on the cusp of formation may contain an additional area fraction $\psi$ 
of one of the components, resulting in a Gaussian rigidity for the pringle of 
the form $\bar{\kappa} + \psi\,\delta\bar{\kappa}$, with $\delta\bar{\kappa} = 
\bar{\kappa}_1 - \bar{\kappa}_2 > 0$ the difference between
the Gaussian rigidities of each component.  The membrane Hamiltonian might then 
include the extra term
\begin{equation}
	\mathcal{H}_\psi = 
	\int_\mathcal{S}\mathrm{d}A\left(\frac{\chi\,\psi^2}{2}\right),
	\label{eq:Hpsi}
\end{equation}
with $\chi$ a Flory-like parameter that approaches zero the bulk membrane 
approaches the demixing transition.

For nanodiscoids subject to a pringle shaped ($\phi = -1$) perturbation, 
\begin{equation}
	\mathcal{H}_\psi = \frac{\chi\,\psi^2\,\pi\,R_0^2}{2}\left( 1 + 
	\frac{\alpha^2}{4} - \frac{\alpha^4}{48} \right) + O\left( \alpha^5 
	\right).
	\label{eq:Hpsi_pert}
\end{equation}
The total energy $\mathcal{H} = \mathcal{H}_m + \mathcal{H}_b + 
\mathcal{H}_\psi$ is then
\begin{equation}
	\begin{split}
	\mathcal{H} &= \pi\bigg\{\left[\frac{k_g}{R_0}\left( 1 - q_0\,R_0 \right) - 
	\frac{R_0^2\,\sigma}{2} + \frac{\chi\,\psi^2\,R_0^2}{2}\right] + 
	\alpha^2\left[ \frac{ \chi\,\psi^2\,R_0^2}{8}-\left( \bar{\kappa} + 
	\psi\,\delta\bar{\kappa} - \bar{\kappa}^\ast_p \right)\right]\\
	&\quad+ \alpha^4\left[ \frac{3}{4}\left(  \bar{\kappa} + 
	\psi\,\delta\bar{\kappa}- \bar{\kappa}^\dagger_p\right) - \frac{ 
	\chi\,\psi^2\,R_0^2}{96} \right]\bigg\} + O\left( \alpha^5 \right).
\end{split}
	\label{eq:H+Hpsi}
\end{equation}
The additional area fraction $\psi$ of the pringle's surface that is occupied 
by the component $\bar{\kappa}_1$ is found by imposing $\partial \mathcal{H} / 
\partial \psi = 0$, which implies
\begin{equation}
	\psi^2 = \frac{\left( \delta\bar{\kappa} \right)^2}{\chi^2\,R_0^4}\alpha^4 
	+ O\left( \alpha^5 \right).
	\label{eq:psi^2}
\end{equation}
Substituting into (\ref{eq:Hpsi_pert}) gives
\begin{equation}
	H_\psi = \frac{\pi \left( \delta\bar{\kappa} \right)^2}{2\,\chi\,R_0^2}+ 
	O\left( \alpha^5 \right),
	\label{eq:Hpsi_pert_subst}
\end{equation}
and hence
\begin{equation}
	\mathcal{H} = \pi\bigg\{\left[\frac{k_g}{R_0}\left( 1 - q_0\,R_0 \right) - 
	\frac{R_0^2\,\sigma}{2}\right] - \alpha^2\left( \bar{\kappa} - 
	\bar{\kappa}^\ast_p \right)
	+\frac{3\alpha^4}{4}\left[  \bar{\kappa} - \left(\bar{\kappa}^\dagger_p - 
	\frac{2\left( \delta\bar{\kappa} \right)^2}{3\,\chi\,R_0^2}\right) 
\right]\bigg\} + O\left( \alpha^5 \right).
\label{eq:H+Hpsi_subst}
\end{equation}
Writing
\begin{equation}
	\bar{\kappa}^\dagger_\chi = \bar{\kappa}^\dagger_p - \frac{2\left( 
		\delta\bar{\kappa} \right)^2}{3\,\chi\,R_0^2},
\end{equation}
implies non-zero stable solutions $\alpha^\ast$, and hence principal curvatures 
of magnitude
\begin{equation}
|c_1| = \frac{1}{R_0}\left[ \frac{2\left( \bar{\kappa} - \bar{\kappa}^\ast_p 
\right)}{3\left( \bar{\kappa} - \bar{\kappa}^\dagger_\chi \right)} 
\right]^{1/2},\label{eq:alphastar_chi}
\end{equation}
for average Gaussian curvatures 
$\bar{\kappa}>\mathrm{max}\left(\bar{\kappa}^\ast_p,\bar{\kappa}^\dagger_\chi\right)$.

\section*{Image analysis}
\begin{figure*}[!t]
	\begin{center}
		\includegraphics[width=0.8\textwidth]{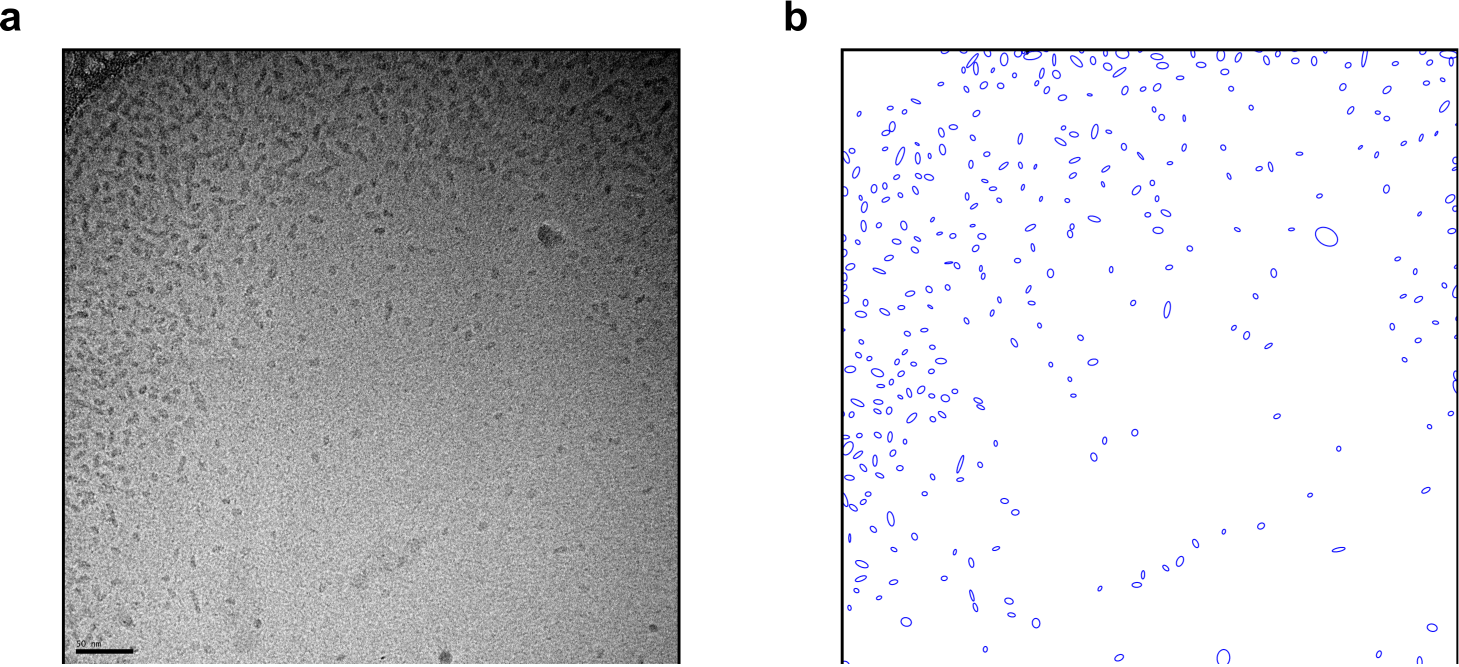}
	\end{center}
	\caption
	{
		(Color online) Cryo Electron Microscopy image ({\bf a}) and associated 
		image analysis ({\bf b}) of Styrene Maleic Acid stabilised 
		nanodiscoids, prepared according to \cite{Lee2016} (scale bar = 50 nm).
	}
\label{fig:img}
\end{figure*}
Two cryo Electron Microscopy images were analysed, one is shown in Fig.~3 of 
the main manuscript, and the other is displayed in Fig.~\ref{fig:img}.  Images 
were analysed according to the following protocol.  In order to remove unwanted 
noise, images were pre-processed using the free application ImageJ.  After 
conversion to an 8-bit single channel image, a small Gaussian blur was applied 
(10 pixel variance) followed by a low pass filter, set to remove structures 
below 15 pixels in size (1 pixel $= 0.131$ nm).  The identification and fitting 
of shapes was carried out using the freely available OpenCV library of Python 
functions.  First, the pre-processed image was thresholded, after which discoid 
contours were extracted.  Ellipses were then fitted to the discoid contours 
using an in-built least-squares procedure.  We remark that the protocol relies 
on edge-detection via contrast, and hence is not well suited to differentiating 
between overlapping SMALPs or identifying those that are out of the focal 
plane.

In total, the analysis identified 414 SMALPs across both images.  The average 
length of the ellipse semi-major axes was 3.48 nm, with a variance of 1.68 
nm$^2$.

\end{widetext}
\end{document}